\documentclass[11pt,a4paper]{article}
\pdfoutput=1
\usepackage{jheppub}

\usepackage{multirow, graphicx,amssymb,url,mathrsfs,amsmath}
\usepackage{wrapfig,boxedminipage,setspace,subfigure,epsfig}
\usepackage{amsxtra,amstext,latexsym,dsfont,amsfonts}
\usepackage{color,eucal}
\usepackage[dvipsnames]{xcolor}
\usepackage{float}
\usepackage{slashed,comment}
\usepackage{kotex}
\usepackage{tikz}
\usetikzlibrary{calc,patterns,angles,quotes}
\usetikzlibrary{decorations.pathreplacing,decorations.markings,snakes}
\usepackage{tabularx, array}
\usepackage{mdframed,mathtools}
\newcolumntype{L}[1]{>{\raggedright\arraybackslash}p{#1}}
\newcolumntype{C}[1]{>{\centering\arraybackslash}p{#1}}
\newcolumntype{R}[1]{>{\raggedleft\arraybackslash}p{#1}}

\newcommand{\be}{\begin{equation}}
\newcommand{\ee}{\end{equation}}
\newcommand{\bea}{\begin{eqnarray}}
\newcommand{\eea}{\end{eqnarray}}

\title{Recursion Coefficients and Krylov Dynamics in Polynomial Random Matrix Models}

\author[a]{Juan F. Pedraza}
\author[a,b]{and Le-Chen Qu}
\emailAdd{j.pedraza@csic.es}
\emailAdd{lechen.qu@ift.csic.es}

\preprint{\texttt{IFT-UAM/CSIC-26-106}}

\affiliation[a]{Instituto de F\'isica Te\'orica UAM/CSIC, Calle Nicol\'as Cabrera 13-15, 28049 Madrid, Spain}
\affiliation[b]{Departamento de F\'isica Te\'orica, Universidad Aut{\'o}noma de Madrid, 28049 Madrid, Spain}

\abstract{We study the recursion coefficients of orthogonal polynomials and their associated Krylov dynamics in random matrix models with high-degree and possibly asymmetric polynomial potentials. We develop a moment recursion method that,
when combined with the recursive algorithm, provides an efficient construction
of the recursion coefficients.  We also obtain their large-$n$
asymptotic behavior for general asymmetric potentials; for $Nw_d=1$, the leading asymptotic form
of $R_n$ reproduces Freud's conjecture.  We apply this framework
to an asymmetric quartic potential and to the double-scaled
Sachdev-Ye-Kitaev (DSSYK) model.  In both models, the recursion functions
capture the overall qualitative behavior of the recursion coefficients, and the
gradient catastrophes of the recursion functions are associated with ``chaotic'' transition regions in the recursion coefficients.
For the quartic potential, such regions can occur in both $R_n$ and $S_n$,
whereas the DSSYK model can exhibit multiple transition regions in $R_n$, with
the recursion function remaining accurate in the smooth intervals between
them.  Finally, we compute the corresponding spread complexity and find that
transition regions do not qualitatively modify its behavior, while a
two branch structure produces early time oscillations followed by monotonic
growth.}

\begin{document}
\maketitle

%

\section{Introduction}

Orthogonal polynomials have long played an important role in both mathematics
and physics.  Their applications to random matrix theory have contributed to
the development of quantum gravity, string theory, and integrable partial
differential equations \cite{demeterfi1993two,witten1993two,ADLER199567}.  More recently, orthogonal polynomials have been
reinterpreted as Krylov polynomials by identifying their recursion coefficients
with Lanczos coefficients \cite{Muck:2022xfc, Kar:2021nbm, Muck:2024fpb, Adhikari:2025vdl, Alishahiha:2024vbf,Balasubramanian:2025xkj,Lunt:2025dcc,Balasubramanian:2022dnj,Qu:2025lgo,Murugan:2026rfa,Qu:2026dmv}.  This connection has renewed interest in the
nonlinear equations governing these coefficients.

These equations have been studied extensively by mathematicians and physicists
over the past fifty years.  In the mathematical literature, they are known as
Freud equations, after G\'eza Freud, who first derived them in the theory of
orthogonal polynomials \cite{freud1976coefficients}.  In the physics literature, they are referred to as
discrete string equations and arise in the enumeration of Feynman graphs in string theory \cite{Brezin:1977sv,Bessis:1980ss,Itzykson:1979fi}.  Although discrete string equations generally do not admit closed form solutions, they provide a direct characterization of
the Lanczos coefficients that complements their recursive construction. 

The recursion coefficients $R_n$ of symmetric random matrix ensembles were
studied extensively in the 1990s, when ``chaotic, pseudo oscillatory'' behavior
was first observed \cite{Jurkiewicz:1991sj,senechal1992chaos}.  Subsequent work has interpreted this ``chaotic phase'' as
a dispersive shock wave in a hydrodynamic chain \cite{Benassi:2019qhd}.  Since the symmetry of the
potential imposes $S_n=0$, previous analyses have focused on $R_n$, and on its large-$n$ asymptotic behavior \cite{deift1999strong,bleher1999semiclassical,baik2003uniform}.  Moreover, the discrete string equations
become increasingly cumbersome as the degree of the potential grows, and most
studies have therefore focused primarily on relatively low degree potentials \cite{demeterfi1990multiband,jurkiewicz1990regularization,lechtenfeld1991phase,sasaki1991matrix,lechtenfeld1992eigenvalue,lechtenfeld1992semiclassical,clarkson2021generalized,clarkson2022symmetric,clarkson2023generalized,clarkson2025symmetric}.

In this paper, we derive the large-$n$ asymptotic behavior of the recursion
coefficients for general asymmetric potentials.  For $Nw_d=1$, the leading
asymptotic form of $R_n$ reproduces Freud's conjecture.  To extend the
computation to high degree polynomial potentials, we combine the
moment recursion method with the recursive algorithm, thereby avoiding the
increasingly cumbersome discrete string equations.

We apply the discrete string equations to the asymmetric quartic potential and
the moment recursion method to the double-scaled
Sachdev-Ye-Kitaev model \cite{Cotler:2016fpe,Garcia-Garcia:2018fns,Berkooz:2018jqr}.  In both cases, the recursion functions capture the overall qualitative behavior of the
recursion coefficients, and gradient catastrophes of the recursion functions
are associated with the appearance of ``chaotic'' transition regions.  For the
quartic potential, such regions can develop in both $R_n$ and $S_n$.  The DSSYK
model can instead exhibit multiple transition regions in $R_n$, while the
recursion function remains accurate in the smooth intervals between successive
transition regions.  Building on Ref.~\cite{Qu:2025lgo}, we also compute the corresponding
spread complexity \cite{Parker:2018yvk,Barbon:2019wsy,Avdoshkin:2019trj,Rabinovici:2020ryf,Jian:2020qpp,Dymarsky:2021bjq,Hornedal:2022pkc,Balasubramanian:2022tpr,He:2022ryk,Caputa:2023vyr,Erdmenger:2023wjg,Craps:2023ivc,Huh:2023jxt,He:2024hkw,He:2024xjp,Caputa:2024vrn,Baggioli:2024wbz,Craps:2024suj,Huh:2024ytz,Caputa:2024sux,Zhai:2024tkz,Nandy:2024mml,Li:2024ljz,Balasubramanian:2024ghv,Bhattacharya:2024szw,Bhattacharya:2024hto,Bhattacharya:2024uxx,Aguilar-Gutierrez:2024nau,Baggioli:2025knt,Craps:2025kub,Evnin:2025cfx,Fu:2025kkh,He:2025guu,Zhai:2025abc,Caputa:2025dep,Caputa:2025ozd,Caputa:2025mii,Miyaji:2025yvm,Takahashi:2025iol,Miyaji:2025ucp,Demulder:2025uda,Alishahiha:2026fnu,Chowdhury:2026fjb,Li:2026jxx,DeRo:2026mlc,Nunez:2026kwr,Das:2026gko,Bhattacharyya:2026zpu,Balasubramanian:2026klv,Caputa:2026ldd} and find that transition regions do not qualitatively alter
its behavior, whereas a two branch structure produces early time oscillations
followed by monotonic growth.

The paper is organized as follows.  Section~\ref{Preliminaries} reviews the
orthogonal polynomial formalism in random matrix theory and its connection to
Krylov dynamics.  In section~\ref{Recursioncoefficientsforpolynomialpotentials},
we develop the moment recursion method and discuss the asymptotic behavior of
the recursion coefficients for general asymmetric potentials.
Sections~\ref{Asymmetricquarticpotential} and~\ref{sec:effective-potential}
analyze the asymmetric quartic potential and the double-scaled SYK model,
respectively.  Finally, section~\ref{Conclusion} summarizes our results and
outlines directions for future work.

\section{Preliminaries}\label{Preliminaries}
This section reviews the orthogonal polynomial formalism in random matrix
theory and its connection to Krylov dynamics \cite{Qu:2025lgo}. For comprehensive reviews, see~\cite{Nandy:2024htc,Baiguera:2025dkc,Rabinovici:2025otw}. Readers familiar with these
topics may skip this review.

\subsection{Random matrix theory}
A random matrix theory is defined by an ensemble of $N\times N$ Hermitian
matrices equipped with the probability
measure~\cite{BESSIS1980109,Ginsparg:1991bi,Eynard:2015aea, bleher2011lectures,Ginsparg:1993is,livan2018introduction}
\begin{equation}
    d\mu_N(H) = \frac{1}{Z_N}\, e^{-N\,\mathrm{Tr}\,V(H)}\, dH ,
\end{equation}
where the potential $V(\lambda)$ is assumed to satisfy
\begin{equation}\label{boundarycondition}
    \lim_{\lambda\to\pm\infty} \bigl( V(\lambda) - \log(\lambda^2+1) \bigr) = +\infty ,
\end{equation}
so that the measure is normalizable.  The Lebesgue measure on the space of
Hermitian matrices is
\begin{equation}
    dH = \prod_{j=1}^N dH_{jj} \prod_{j<k} d\,\mathrm{Re}\,H_{jk}\, d\,\mathrm{Im}\,H_{jk},
\end{equation}
and the normalization factor, or partition function, is
\begin{equation}
    Z_N = \int e^{-N\,\mathrm{Tr}\,V(H)}\, dH.
\end{equation}
The measure is invariant under unitary conjugation,
\begin{equation}
    H \;\longrightarrow\; U^{-1}HU , \qquad U\in U(N),
\end{equation}
and the Weyl integration formula therefore gives the following joint
probability density for the eigenvalues of $H$:
\begin{equation}
    d\mu_N(\lambda)
    = \frac{1}{N!\Delta_N}\,
      \prod_{j>k}(\lambda_j - \lambda_k)^2
      \exp\!\left[-N\sum_{i=1}^N V(\lambda_i)\right]
      \prod_{i=1}^N d\lambda_i ,
\end{equation}
where the normalization constant is
\begin{equation}\label{normalcon}
    \Delta_N
    = \frac{1}{N!}\int
      \prod_{j>k}(\lambda_j - \lambda_k)^2
      \exp\!\left[-N\sum_{i=1}^N V(\lambda_i)\right]
      \prod_{i=1}^N d\lambda_i .
\end{equation}

\subsection{Orthogonal polynomials}
Random matrix theory can be analyzed efficiently through the associated monic
orthogonal polynomials
\(
P_n(\lambda) = \lambda^{n} + \text{(lower-order terms)},
\)
defined with respect to the inner product~\cite{BESSIS1980109,Ginsparg:1991bi,Eynard:2015aea, bleher2011lectures,Ginsparg:1993is,livan2018introduction}
\begin{equation}\label{innerproduct}
    \int_{-\infty}^{\infty}
    P_n(\lambda)\, P_m(\lambda)\,
    e^{-N V(\lambda)}\, d\lambda
    = h_n\, \delta_{mn}\, .
\end{equation}
The constants $h_n$ are the squared norms, with
\(
h_0 = \int_{-\infty}^{\infty} d\mu(\lambda),
\)
and
\(
d\mu(\lambda) = e^{-N V(\lambda)}\, d\lambda
\).
Heine's formula gives the $n$-th monic polynomial as the ensemble average of a
characteristic determinant,
\begin{equation}\label{Heineformula}
P_n(\lambda)
= \mathbb{E}\!\left[\det(\lambda I_n - H)\right]
=
\frac{1}{n!\Delta_n}\int
\prod_{i=1}^{n} (\lambda - \lambda_i)\,
\prod_{i<j} |\lambda_i - \lambda_j|^{2}\,
e^{-N \sum_{i=1}^{n} V(\lambda_i)}\,
d\lambda_1 \cdots d\lambda_n
 .
\end{equation}
Here $\mathbb{E}[\cdots]$ denotes the expectation with respect to the
joint eigenvalue distribution of $H$, and $I_n$ is the identity matrix.  All correlation functions are generated by the
Christoffel-Darboux kernel
\begin{equation}\label{Dysondeterminantalormula}
\begin{aligned}
K_N(\lambda_i,\lambda_j)
&=
\sum_{n=0}^{N-1} \frac{1}{h_n}\,
P_n(\lambda_i)\, P_n(\lambda_j)\,
e^{-\frac{N}{2} V(\lambda_i)}
e^{-\frac{N}{2} V(\lambda_j)}.
\end{aligned}
\end{equation}
The monic orthogonal polynomials satisfy the three term recurrence relation
\begin{equation}\label{recursionrelation}
    \lambda\, P_n(\lambda)
    = P_{n+1}(\lambda)
      + S_n\, P_n(\lambda)
      + R_n\, P_{n-1}(\lambda)\, ,
\end{equation}
with $P_{-1}=0$ and $P_0=1$.  The recursion coefficients $R_n$ and $S_n$ are determined by
\begin{equation}
R_n
=\frac{1}{h_{n-1}}
\int_{-\infty}^{\infty}
\lambda P_{n-1}(\lambda)P_n(\lambda)d\mu(\lambda),
\qquad
S_n
=
\frac{1}{h_n}
\int_{-\infty}^{\infty}
\lambda P_n^2(\lambda)d\mu(\lambda).
\end{equation}
Relations among these coefficients follow from the identities
\begin{equation}\label{twoidentitiess}
\begin{aligned}
    \int d\mu(\lambda)\, P_n(\lambda)\, \frac{d}{d\lambda} P_n(\lambda) &= 0,\\
    \int d\mu(\lambda)\, P_{n-1}(\lambda)\, \frac{d}{d\lambda} P_n(\lambda) &= n\, h_{n-1}.
\end{aligned}
\end{equation}
It is useful to introduce an auxiliary Hilbert space spanned by the normalized
states
\begin{equation}
    |n\rangle = \frac{P_n(\lambda)}{\sqrt{h_n}}, \qquad n = 0,1,2,\ldots ,
\end{equation}
which form an orthonormal basis,
\begin{equation}\label{orthonormalbasis}
    \langle n | m \rangle = \delta_{mn}.
\end{equation}
In this basis, the recurrent relation \eqref{recursionrelation} becomes
\begin{equation}\label{compactrecursionrelation}
    \hat{\lambda}\, |n\rangle
    = \sqrt{R_{n+1}}\, |n+1\rangle
      + S_n\, |n\rangle
      + \sqrt{R_n}\, |n-1\rangle,
\end{equation}
Using the normalizability condition \eqref{boundarycondition}, the identities
\eqref{twoidentitiess} can therefore be written as the discrete string
equations
\begin{equation}\label{VSRREEQ}
\begin{aligned}
\langle n | V'(\hat{\lambda}) | n \rangle &= 0 ,\\
\sqrt{R_n}\,\langle n-1 | V'(\hat{\lambda}) | n \rangle &= \frac{n}{N} .
\end{aligned}
\end{equation}
In the large $N$ limit, with $x=n/N$, these equations reduce to the continuum
string equations \cite{Balasubramanian:2022dnj,Qu:2025lgo}\footnote{Throughout this work, the term ``string equations''
refers to the discrete string equations in the double-scaling limit, whereas
their large $N$ limits are referred to as ``continuum string equations''.}
\begin{equation}\label{extremaofthepotential}
\begin{aligned}
2x
&=
\sqrt{R(x)}
\frac{\partial}{\partial\sqrt{R(x)}}
\left[
\int d\lambda\,
\frac{
V(\lambda)
}{
\pi\sqrt{4R(x)-(\lambda-S(x))^2}
}
\right],
\\[4pt]
0
&=
\frac{\partial}{\partial S(x)}
\left[
\int d\lambda\,
\frac{
V(\lambda)
}{
\pi\sqrt{4R(x)-(\lambda-S(x))^2}
}
\right].
\end{aligned}
\end{equation}
For a polynomial potential
\(
    V(\lambda)=\sum_{m=0}^{d}w_m\lambda^m,
\)
the integrals in \eqref{extremaofthepotential} can be evaluated explicitly.
The continuum string equations then become algebraic relations for the
recursion functions $R(x)$ and $S(x)$:
\begin{subequations}\label{polyabv}
\begin{align}
    0
    &= \sum_{m=0}^{d} w_m
       \sum_{q=0}^{m-1}
       (m-q)\, S(x)^{\,m-q-1} R(x)^{\,q/2}
       \binom{m}{q}\binom{q}{q/2},
       \label{polyabv-S}
    \\
    x 
    &= \frac{1}{2}
       \sum_{m=0}^{d} w_m
       \sum_{q=2}^{m}
       q\, S(x)^{\,m-q} R(x)^{\,q/2}
       \binom{m}{q}\binom{q}{q/2}.
       \label{polyabv-R}
\end{align}
\end{subequations}
Remarkably, for $0<x<1$, Eq.~\eqref{polyabv} also describes the average
Lanczos coefficients upon reversing the continuum coordinate, $x\to1-x$.

\subsection{Orthogonal polynomials as Krylov polynomials}
The recurrent relation \eqref{compactrecursionrelation} is precisely the
tridiagonal form of a Hamiltonian in the Krylov basis.  Identifying
\(\hat{\lambda}\) with the Hamiltonian, the time evolution from the initial
state \(|0\rangle\) is given by \cite{Qu:2025lgo}
\begin{equation}\label{krylovdyna}
|\psi(t)\rangle
= e^{-i\hat{\lambda} t} |0\rangle
= \sum_{n=0}^{\infty} \frac{(-i t)^{n}}{n!}\, \hat{\lambda}^{\,n} |0\rangle
= \sum_{n=0}^{\infty} \psi_n(t)\, |n\rangle .
\end{equation}
The Krylov amplitudes obey the discrete Schr\"odinger equation
\begin{equation}\label{psieq}
i\,\partial_t \psi_n(t)
= b_{n+1}\,\psi_{n+1}(t)
 + a_n\,\psi_n(t)
 + b_n\,\psi_{n-1}(t),
\qquad 
\psi_n(0)=\delta_{n0},
\end{equation}
where the Lanczos coefficients are identified with the recursion coefficients,
\(
a_n=S_n
\)
and
\(
b_n=\sqrt{R_n}
\).
Thus the orthonormal polynomials associated with \(d\mu(\lambda)\) provide a
natural realization of Krylov polynomials
\cite{Muck:2022xfc, Kar:2021nbm, Muck:2024fpb, Adhikari:2025vdl, Alishahiha:2024vbf,Balasubramanian:2025xkj,Lunt:2025dcc,Balasubramanian:2022dnj,Qu:2025lgo}.
This perspective also allows the recursion coefficients to be reconstructed
directly from the moments of the measure,
\begin{equation}\label{moments}
\mu_k
= \int \lambda^k\, d\mu(\lambda),
\end{equation}
via the recursive algorithm \cite{viswanath1994recursion,bhattacharjee2023operator}
\begin{equation}\label{recursivealgorithm}
\begin{aligned}
M_k^{(0)}
&=
(-1)^k \frac{\mu_k}{\mu_0},
\qquad
L_k^{(0)}
=
(-1)^{k+1} \frac{\mu_{k+1}}{\mu_0},
\\
M_k^{(n)}
&=
L_k^{(n-1)}
-
L_{n-1}^{(n-1)}
\frac{M_k^{(n-1)}}{M_{n-1}^{(n-1)}}, \qquad
L_k^{(n)}
=
\frac{M_{k+1}^{(n)}}{M_n^{(n)}}
-
\frac{M_k^{(n-1)}}{M_{n-1}^{(n-1)}},
\qquad
k\ge n,
\\
R_n
&=
M_n^{(n)},
\qquad
S_n
=
-
L_n^{(n)}.
\end{aligned}
\end{equation}
Finally, the spread of the state in Krylov space is quantified by the spread
complexity \cite{Balasubramanian:2022tpr},
\begin{equation}
C(t)
= \sum_n n\, |\psi_n(t)|^2
= \sum_n \frac{n}{h_0 h_n}
  \int d\mu(\lambda)\, d\mu(\lambda')\,
P_n(\lambda)\, P_n(\lambda')\, 
e^{-i(\lambda - \lambda')t},
\end{equation}
which measures the growth of the state's support in the Krylov basis, or
equivalently the effective spread of complexity over time.

\section{Recursion coefficients for polynomial potentials}\label{Recursioncoefficientsforpolynomialpotentials}
In this section, we develop the moment recursion method and discuss the asymptotic behavior of
the recursion coefficients for general asymmetric potentials.

\subsection{Moment recursion method}\label{Recursiveconstructionfrommoments}
In previous work \cite{Qu:2025lgo}, the recursion coefficients and spread complexity were obtained
analytically for quadratic potentials, whereas closed form results are generally
unavailable for general potentials.  In such cases, the recursion
coefficients can be determined by treating the discrete string equation
\eqref{VSRREEQ} as an initial value problem, an approach commonly referred to as
the ``orthogonal polynomial method'' \cite{demeterfi1990multiband,jurkiewicz1990regularization,lechtenfeld1991phase,sasaki1991matrix,lechtenfeld1992eigenvalue,lechtenfeld1992semiclassical,clarkson2021generalized,clarkson2022symmetric,clarkson2023generalized,clarkson2025symmetric}.  In the next section, this method is
applied to the asymmetric quartic potential.  For higher degree potentials, such as that
of the DSSYK model, the discrete string equations become cumbersome, and it is
more efficient to compute the recursion coefficients directly from the
moments.  The
key observation is that, for a polynomial potential
\begin{equation}
V(\lambda)=\sum_{m=0}^{d}w_m\lambda^m,
\qquad
\mu_n=\int d\lambda\,\lambda^n e^{-NV(\lambda)} ,
\end{equation}
the moments obey a closed recursion relation.  This follows from the
total derivative identity
\begin{equation}
0
=
\int d\lambda\,
\frac{d}{d\lambda}
\left(\lambda^n e^{-NV(\lambda)}\right)
=
n\mu_{n-1}
-
N
\int d\lambda\,\lambda^n V'(\lambda)e^{-NV(\lambda)},
\end{equation}
where the boundary contribution vanishes by the normalizability condition
\eqref{boundarycondition}.  Substituting
$V'(\lambda)=\sum_{m=1}^{d}m w_m\lambda^{m-1}$ into this identity and solving
for the highest moment gives
\begin{equation}\label{recursivigheent}
\mu_{n+d-1}
=
\frac{
\frac{n}{N}\mu_{n-1}
-
\sum_{m=1}^{d-1}
mw_m\mu_{n+m-1}
}{
dw_d
},
\qquad
n=0,1,2,\ldots.
\end{equation}
Thus, given the initial moments
\begin{equation}
\mu_0,\mu_1,\ldots,\mu_{d-2}
\end{equation}
all higher moments can be generated iteratively.  The recursion coefficients
then follow from the recursive algorithm \eqref{recursivealgorithm}.  Since
this procedure strongly amplifies errors in the input moments, the initial
moments must be evaluated at high precision.  A similar numerical sensitivity
arises when the recursion coefficients are computed by the discrete
string equations, as discussed in Ref.~\cite{clarkson2025symmetric}.

\subsection{Large-$n$ asymptotics}\label{asymptotics}
Although the full sequences of recursion coefficients generally require a
numerical treatment, their large-$n$ behavior can be derived analytically.  We
assume that $R_n$ and $S_n$ approach the recursion functions $R(x)$ and $S(x)$
in the sense that
\begin{equation}\label{assum}
\lim_{n\to\infty}
\frac{R_n}{R(x)}
=
1,
\qquad
\lim_{n\to\infty}
\frac{S_n}{S(x)}
=
1,
\end{equation}
where $x=n/N$.  We further assume that $R(x)$ diverges while $S(x)$ remains
bounded as $x\to\infty$.  Under these assumptions, the dominant contribution
to Eq.~\eqref{polyabv-R} arises from the highest degree monomial
$w_d\lambda^d$, corresponding to $m=d$ and $q=d$, and gives
\begin{equation}
x
=
\frac{d}{2}
\binom{d}{d/2}
w_d
R(x)^{d/2}
+
O\!\left(R(x)^{(d-1)/2}\right).
\end{equation}
Consequently,
\begin{equation}
\lim_{x\to\infty}
\frac{R(x)}{x^{2/d}}
=
\left[
\frac{d}{2}
\binom{d}{d/2}
w_d
\right]^{-2/d}
=
\left[
\frac{
w_d\Gamma(d+1)
}{
\Gamma(d/2)\Gamma(d/2+1)
}
\right]^{-2/d},
\end{equation}
where the second equality follows from
\begin{equation}
\frac{d}{2}
\binom{d}{d/2}
=
\frac{\Gamma(d+1)}
{\Gamma(d/2)\Gamma(d/2+1)}.
\end{equation}
At leading order $R(x)^{(d-2)/2}$, Eq.~\eqref{polyabv-S} receives
contributions from the two highest degree monomials.  The first contribution from
$w_d\lambda^d$, obtained for $m=d$ and $q=d-2$, is
\begin{align}
&
w_d
\bigl(d-(d-2)\bigr)
S(x)^{d-(d-2)-1}
R(x)^{(d-2)/2}
\binom{d}{d-2}
\binom{d-2}{(d-2)/2}
\notag\\
&=
d(d-1)
w_d
S(x)
R(x)^{(d-2)/2}
\binom{d-2}{(d-2)/2}.
\end{align}
The second contribution from $w_{d-1}\lambda^{d-1}$, obtained for $m=d-1$ and
$q=d-2$, is
\begin{align}
&
w_{d-1}
\bigl((d-1)-(d-2)\bigr)
S(x)^{(d-1)-(d-2)-1}
R(x)^{(d-2)/2}
\binom{d-1}{d-2}
\binom{d-2}{(d-2)/2}
\notag\\
&=
(d-1)
w_{d-1}
R(x)^{(d-2)/2}
\binom{d-2}{(d-2)/2}.
\end{align}
Retaining these leading contributions, Eq.~\eqref{polyabv-S} reduces to
\begin{equation}
0
=
(d-1)
R(x)^{(d-2)/2}
\binom{d-2}{(d-2)/2}
\left(
dw_dS(x)+w_{d-1}
\right)
+
O\!\left(R(x)^{(d-4)/2}\right),
\end{equation}
and therefore
\begin{equation}
\label{eq:S-large-x}
\lim_{x\to\infty}
S(x)
=
-\frac{w_{d-1}}{dw_d}.
\end{equation}
Returning to $x=n/N$, define the leading asymptotic quantities
\begin{equation}\label{eq:lanczos-asymptotics}
\begin{aligned}
R_\infty(n)=
n^{2/d}
\left[
\frac{
Nw_d\Gamma(d+1)
}{
\Gamma(d/2)\Gamma(d/2+1)
}
\right]^{-2/d},\qquad
S_\infty=
-\frac{w_{d-1}}{dw_d}.
\end{aligned}
\end{equation}
Combining these expressions with the continuum assumption \eqref{assum} gives
\begin{equation}
\lim_{n\to\infty}
\frac{R_n}{R_\infty(n)}
=1,
\qquad
\lim_{n\to\infty}
\frac{S_n}{S_\infty}
=1.
\end{equation}
Here $R_\infty(n)$ denotes the leading large-$n$ asymptotic form of $R_n$, whereas
$S_\infty$ denotes the limiting value of $S_n$.  Upon imposing the normalization
$Nw_d=1$, the resulting asymptotic formula for $R_n$ coincides precisely with
Freud's conjecture for the recursion
coefficients~\cite{freud1976coefficients,magnus1986freud,lubinsky1988uniform,lubinsky1988proof}. See Ref.~\cite{Murugan:2026rfa} for recent developments on subleading corrections to Freud-type asymptotics and their interpretation in the SYK model. Consequently, the leading
asymptotics of $R_n$ are governed entirely by the highest degree monomial of the
potential, whereas the limit of $S_n$ is fixed by the ratio $w_{d-1}/w_d$.

\section{Asymmetric quartic potential}\label{Asymmetricquarticpotential}
This section presents a numerical study of the recursion coefficients $R_n$
and $S_n$ for an asymmetric quartic potential.  Previous studies have examined
a broad class of symmetric potentials, including symmetric quartic potentials,
for which $S_n$ vanishes identically \cite{demeterfi1990multiband,jurkiewicz1990regularization,lechtenfeld1991phase,sasaki1991matrix,lechtenfeld1992eigenvalue,lechtenfeld1992semiclassical,clarkson2021generalized,clarkson2022symmetric,clarkson2023generalized,clarkson2025symmetric}.  To extend the analysis to asymmetric
potentials, we consider
\begin{equation}
V_q(\lambda)
=
\lambda^4+w_3\lambda^3+w_2\lambda^2+w_1\lambda.
\end{equation}
Substitution into Eq.~\eqref{VSRREEQ} yields the coupled discrete string
equations for $R_n$ and $S_n$,
\begin{equation}\label{discretestringequationsq}
\begin{aligned}
0
={}&
4\left(S_n^3+R_nS_{n-1}+2R_nS_n+2R_{n+1}S_n
+R_{n+1}S_{n+1}\right)
\\
&+3w_3\left(R_n+R_{n+1}+S_n^2\right)+2w_2S_n+w_1,
\\[3pt]
\frac{n}{N}
={}&
4R_n\left(R_{n-1}+R_n+R_{n+1}+S_{n-1}^2+S_{n-1}S_n+S_n^2\right)
\\
&+3w_3R_n(S_{n-1}+S_n)+2w_2R_n.
\end{aligned}
\end{equation}
As mentioned above, the discrete string equations \eqref{discretestringequationsq} can be recast as a forward iteration once the initial conditions are specified.  Solving them
for $R_{n+1}$ and $S_{n+1}$ yields
\begin{equation}\label{quartic-iteration}
\begin{aligned}
R_{n+1}
={}&
\frac{n}{4NR_n}
-R_{n-1}-R_n
-S_{n-1}^2-S_{n-1}S_n-S_n^2
\\
&
-\frac{3w_3}{4}(S_{n-1}+S_n)
-\frac{w_2}{2},
\\[3pt]
S_{n+1}
={}&
-2S_n-\frac{3w_3}{4}
-\frac{1}{4R_{n+1}}\Bigl(
4S_n^3
+4R_nS_{n-1}
+8R_nS_n
\\
&\hspace{8em}
+3w_3R_n
+3w_3S_n^2
+2w_2S_n
+w_1
\Bigr).
\end{aligned}
\end{equation}
The initial conditions are
\begin{equation}
  R_0=0,\quad S_0=\frac{\mu_1}{\mu_0}
 ,\quad R_1=\frac{\mu_0\mu_2-\mu_1^2}{\mu_0^2},
  \quad
  S_1=
  \frac{
    \mu_0^2\mu_3-2\mu_0\mu_1\mu_2+\mu_1^3
  }{
    \mu_0(\mu_0\mu_2-\mu_1^2)
  },
\end{equation}
Here $\mu_n$ denotes the $n$th moment,
\begin{equation}\label{momentvq}
\mu_n
=
\int d\lambda\,\lambda^n e^{-NV_q(\lambda)}.
\end{equation}
With these initial conditions, Eq.~\eqref{quartic-iteration} determines both
$R_n$ and $S_n$ iteratively for all $n\geq2$.
\begin{figure}[!htbp]
\centering
\subfigure[$N=400$]{
  \includegraphics[width=0.31\textwidth]{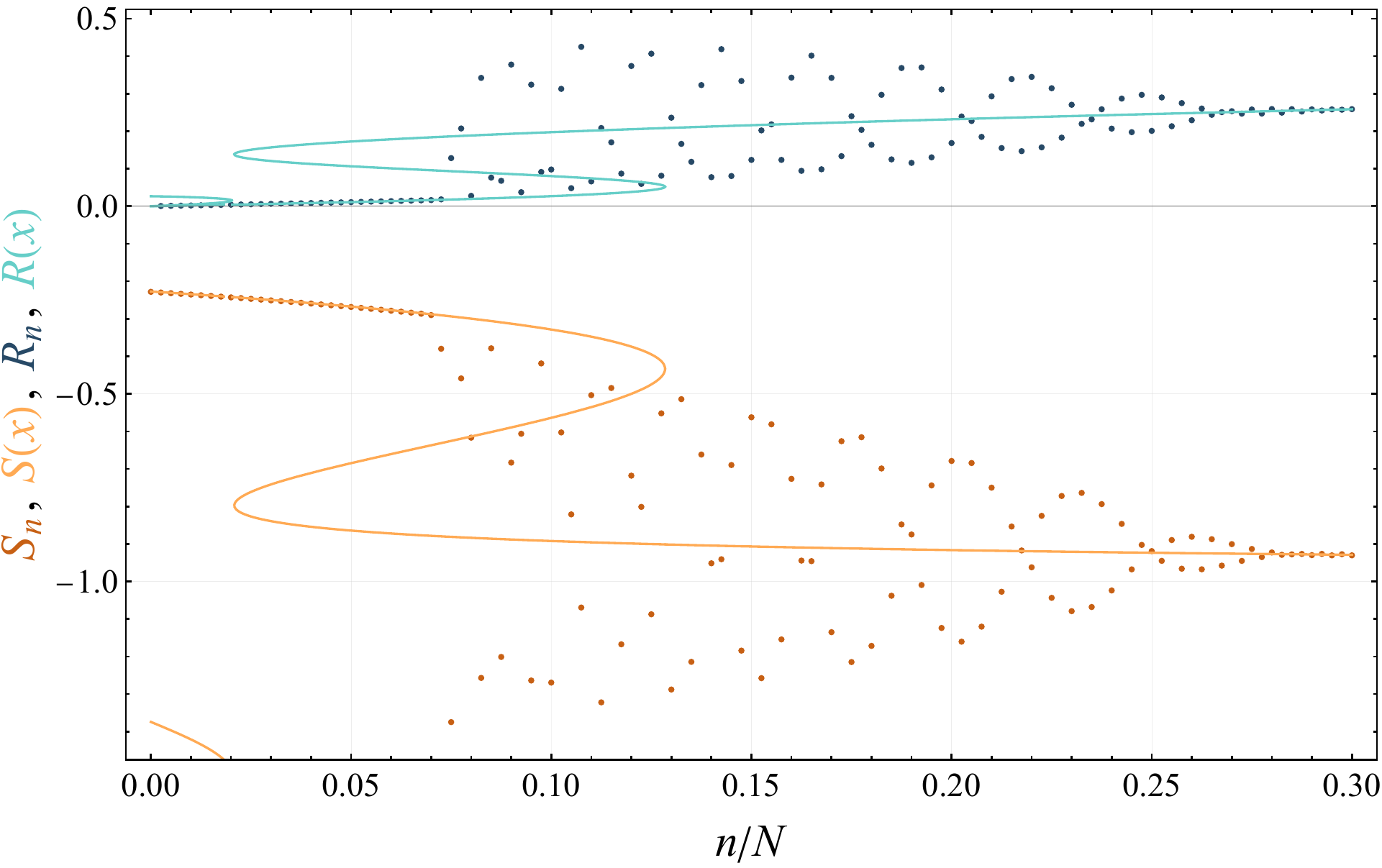}}
\hfill
\subfigure[$N=600$]{
  \includegraphics[width=0.31\textwidth]{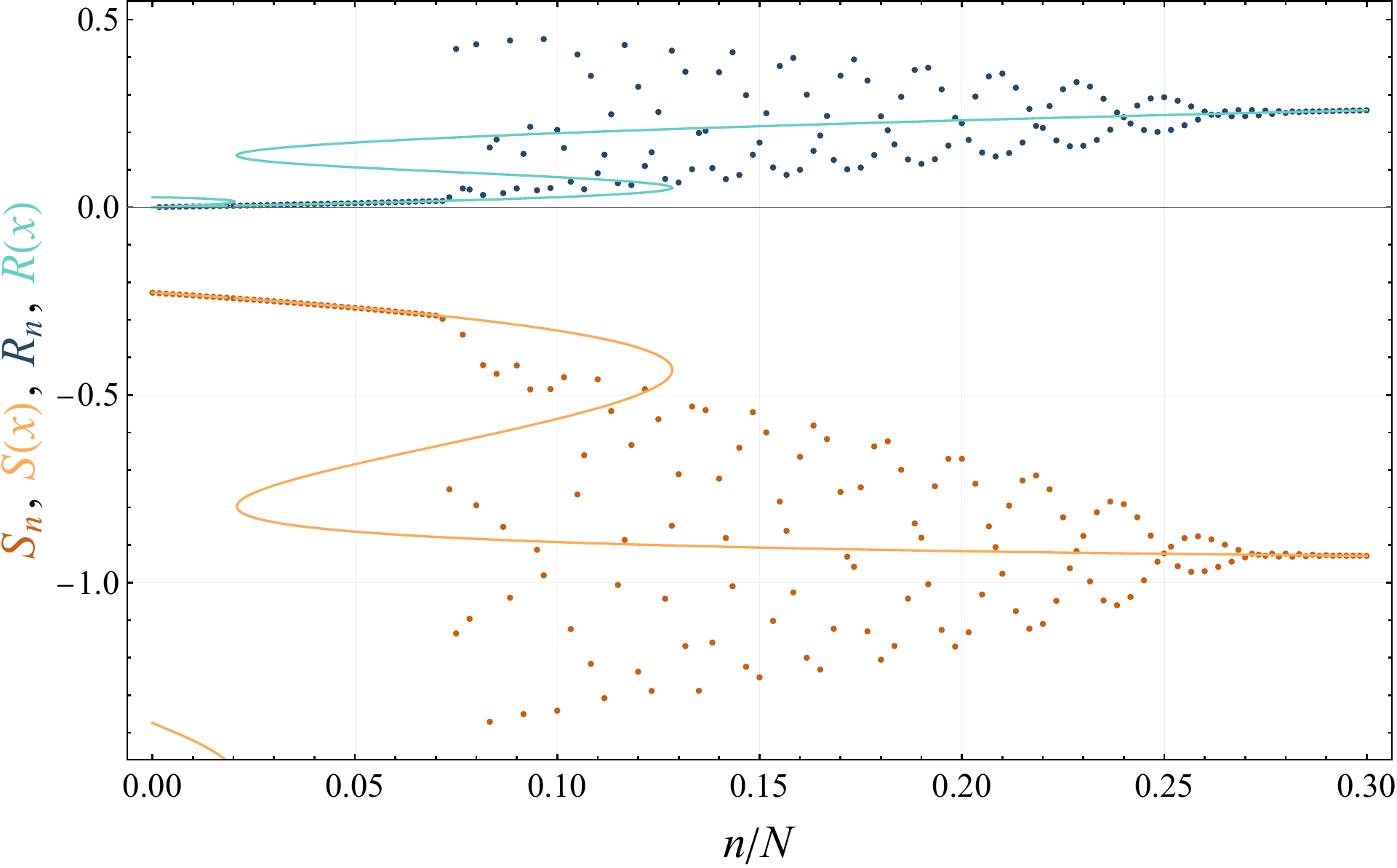}}
\hfill
\subfigure[$N=800$]{
  \includegraphics[width=0.31\textwidth]{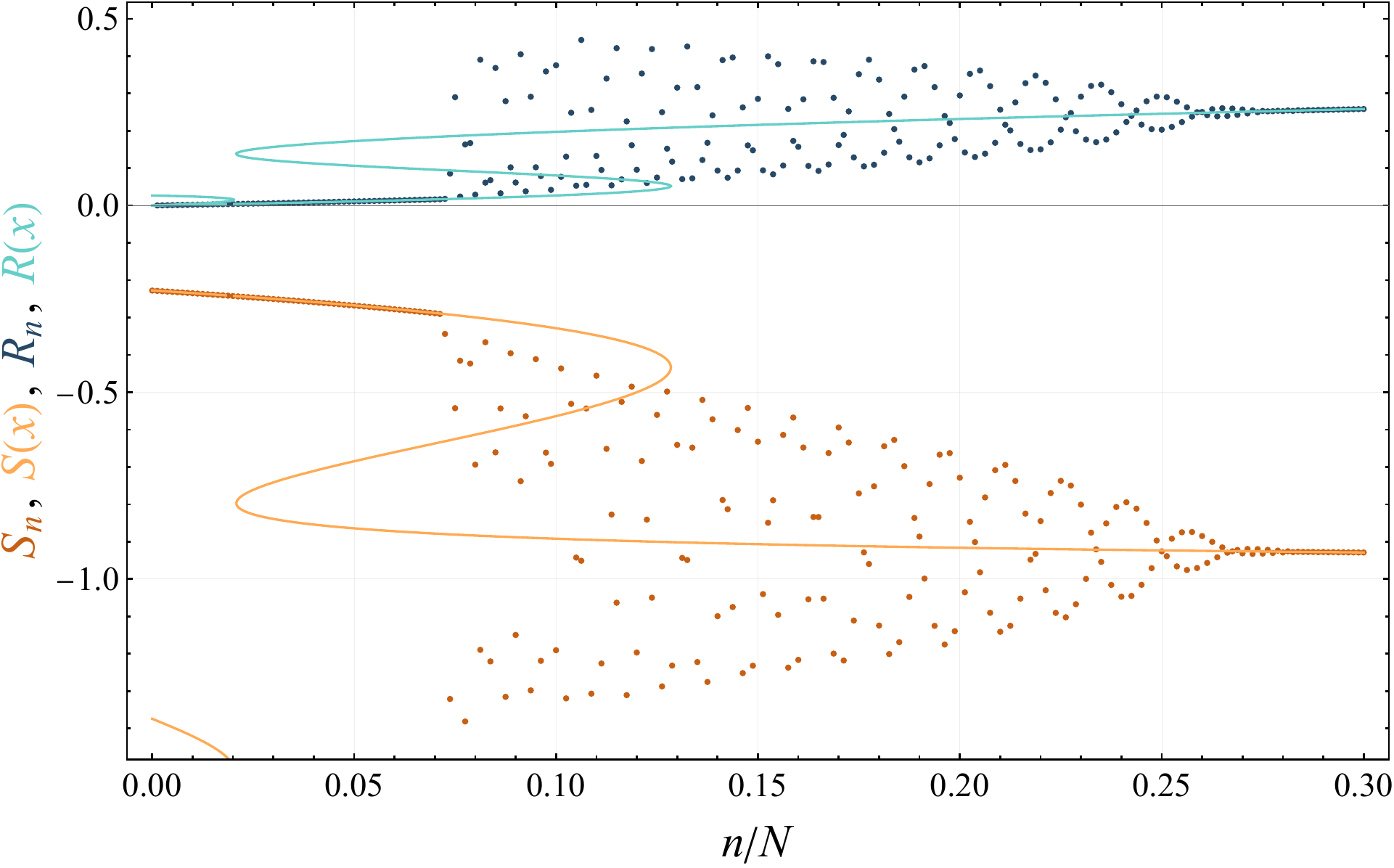}}
\caption{Plots of the recursion coefficients $R_n$ and $S_n$
with the recursion functions $R(x)$ and $S(x)$ for $w_3=4$,
$\alpha=-1$ and $\beta=-0.3$ at $N=400$, $600$ and $800$.}
\label{invariantundertransition}
\end{figure}
The initial moments $\mu_0$, $\mu_1$, and
$\mu_2$ are evaluated by Eq.~\eqref{momentvq} numerically, while $\mu_3$ is
obtained using the moment recursion method based on
Eq.~\eqref{recursivigheent}.
\begin{figure}[!htbp]
\centering

\subfigure[$w_3=4, \alpha=-1, \beta=1.5$]{
  \includegraphics[width=0.31\textwidth]{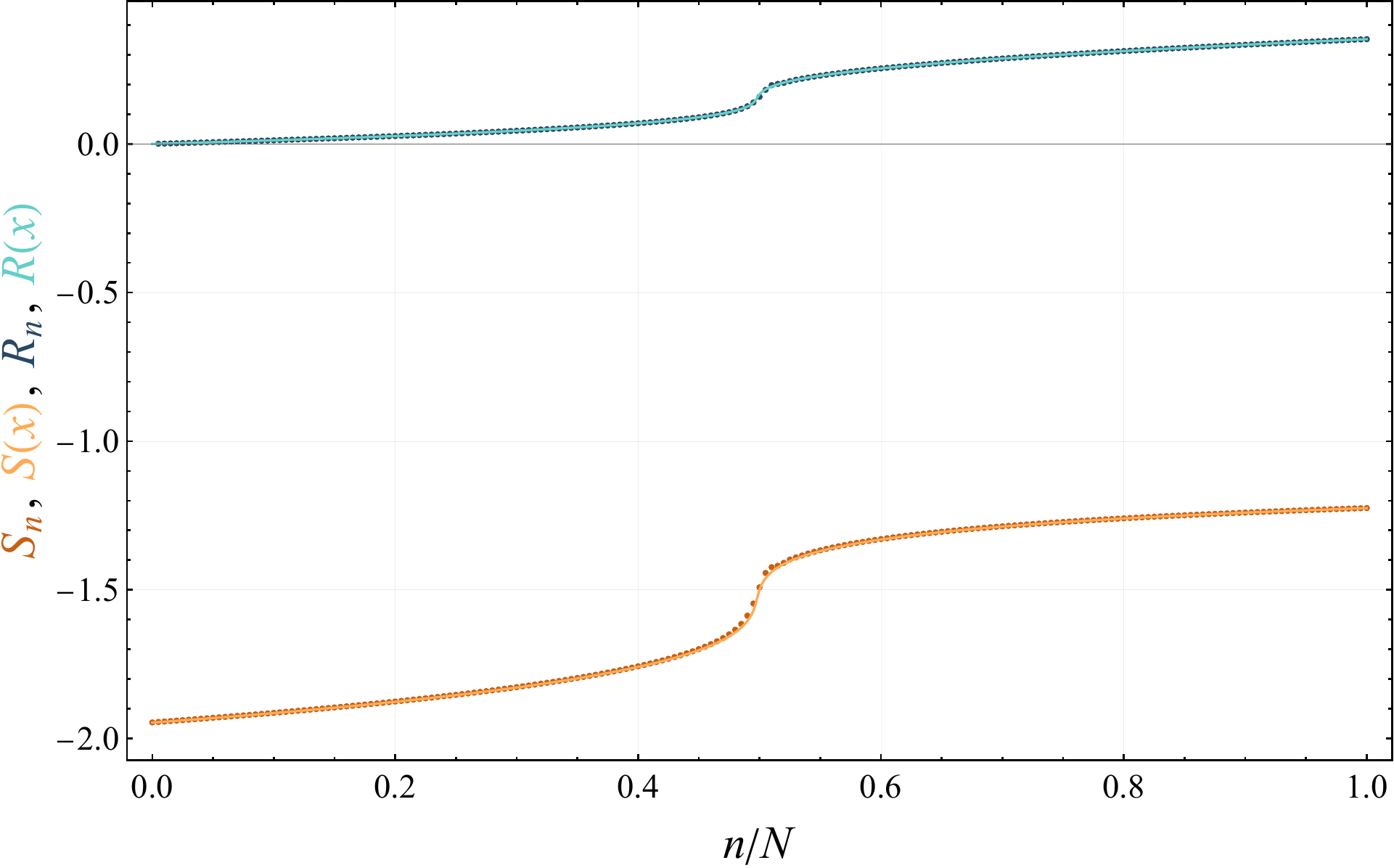}}
\hfill
\subfigure[$w_3=4, \alpha=-1, \beta=\sqrt{-\gamma_2}$]{
  \includegraphics[width=0.31\textwidth]{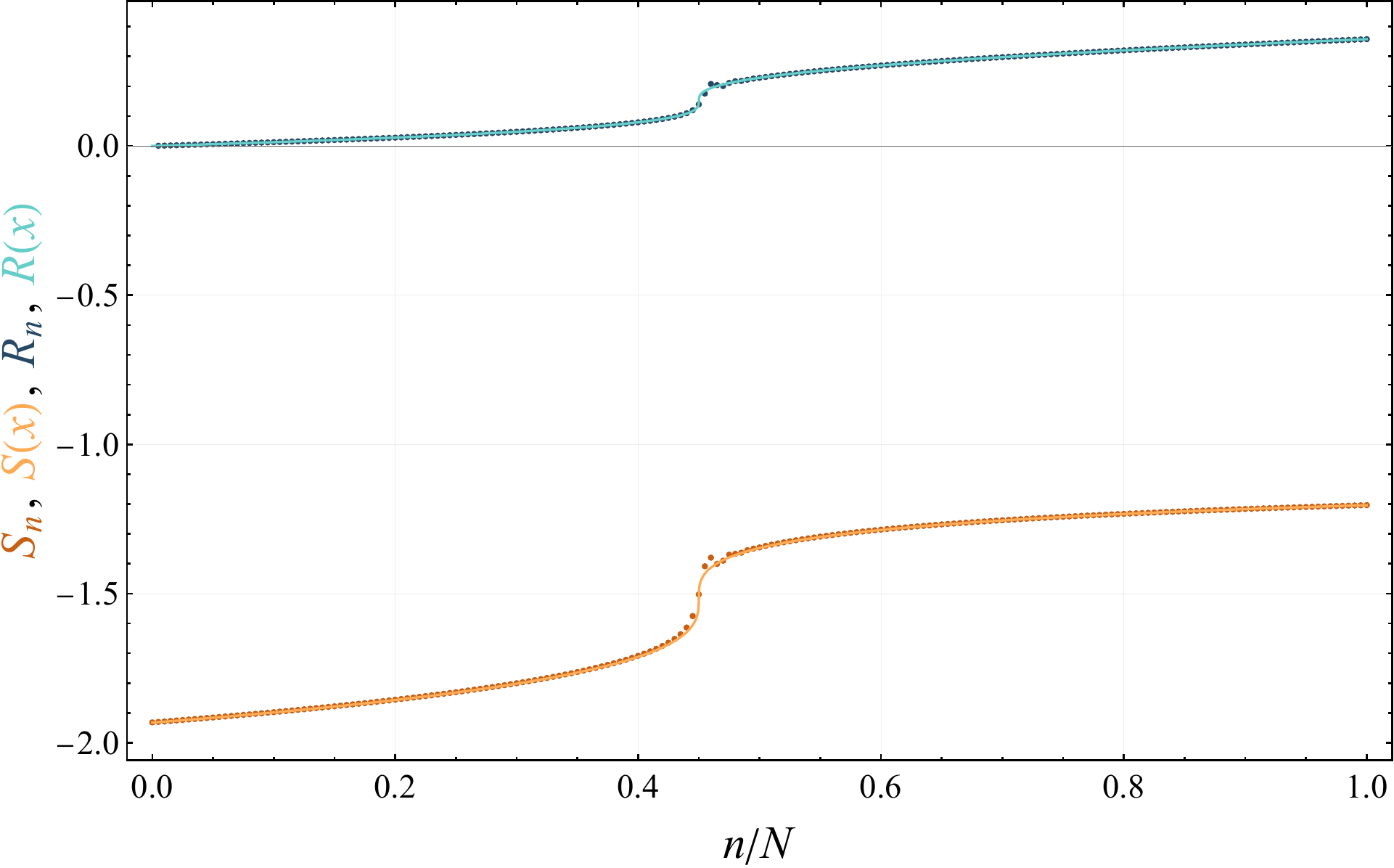}}
\hfill
\subfigure[$w_3=4, \alpha=-1, \beta=1$]{
  \includegraphics[width=0.31\textwidth]{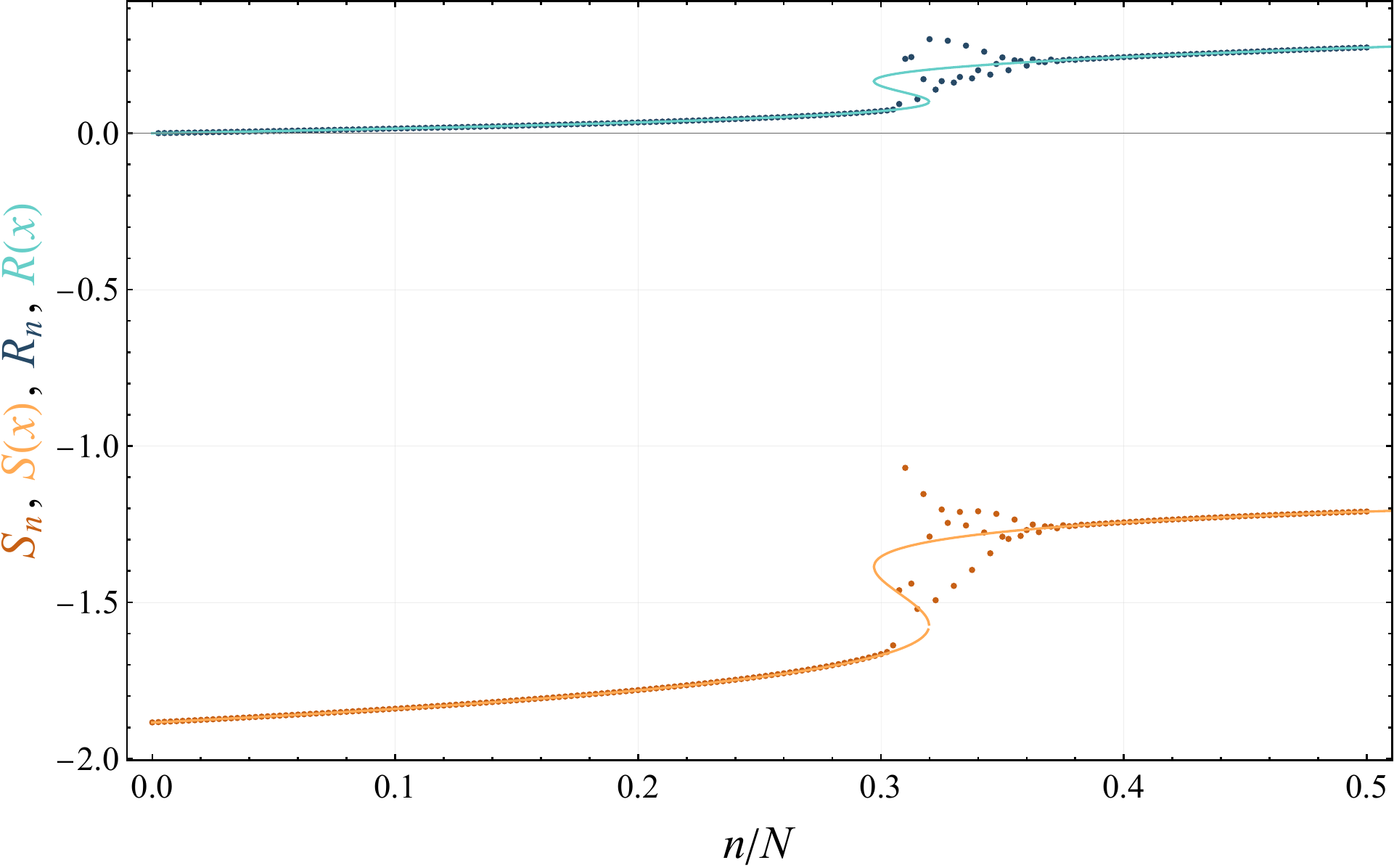}}

\par\medskip

\subfigure[$w_3=4, \alpha=-1, \beta=-1.5$]{
  \includegraphics[width=0.31\textwidth]{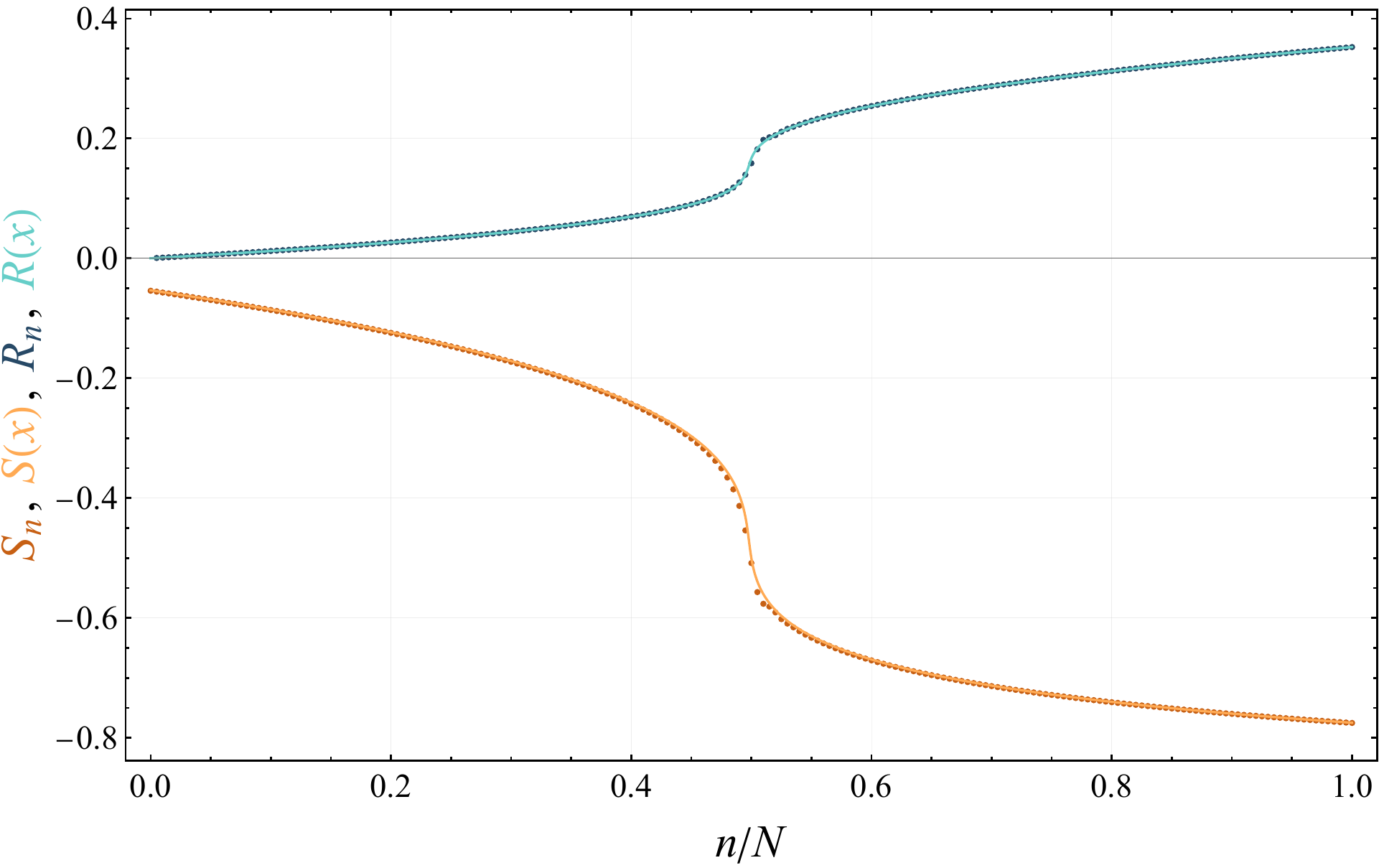}}
\hfill
\subfigure[$w_3=4, \alpha=-1, \beta=-\sqrt{-\gamma_2}$]{
  \includegraphics[width=0.31\textwidth]{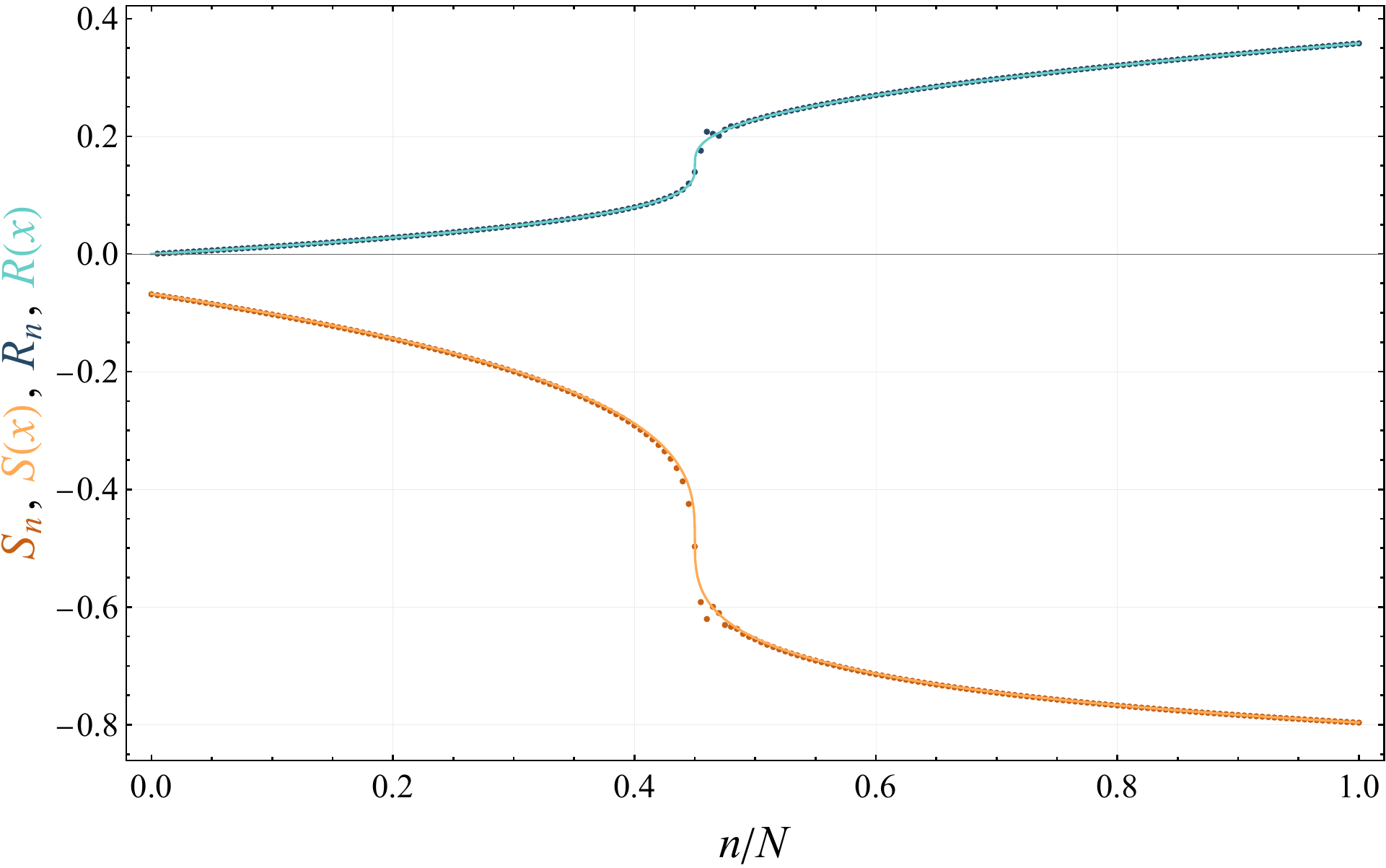}}
\hfill
\subfigure[$w_3=4, \alpha=-1, \beta=-1$]{
  \includegraphics[width=0.31\textwidth]{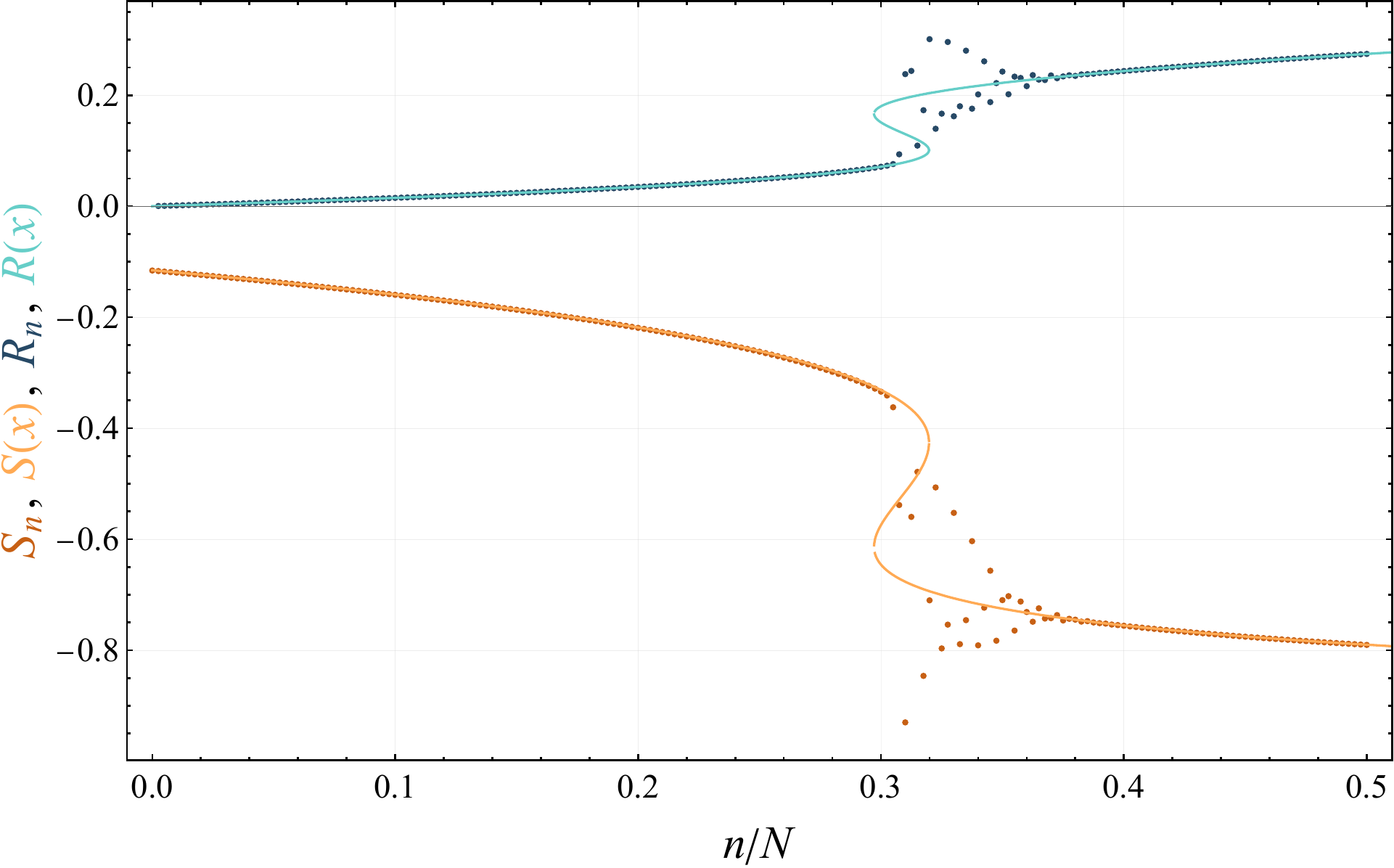}}

\par\medskip

\subfigure[$w_3=4, \alpha=-1, \beta=\sqrt{-\gamma_1}$]{
  \includegraphics[width=0.31\textwidth]{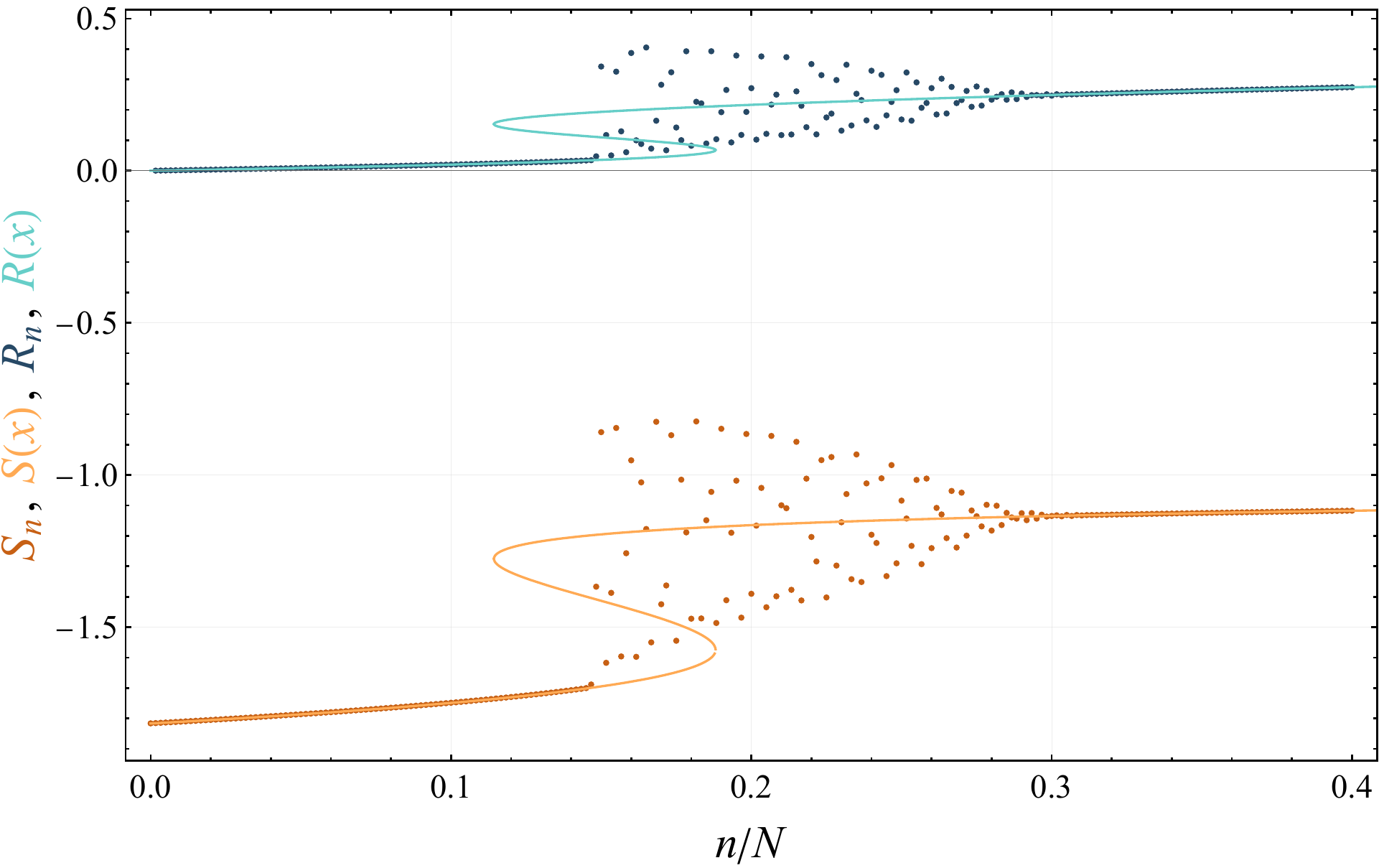}}
\hfill
\subfigure[$w_3=4, \alpha=-1, \beta=0.3$]{
  \includegraphics[width=0.31\textwidth]{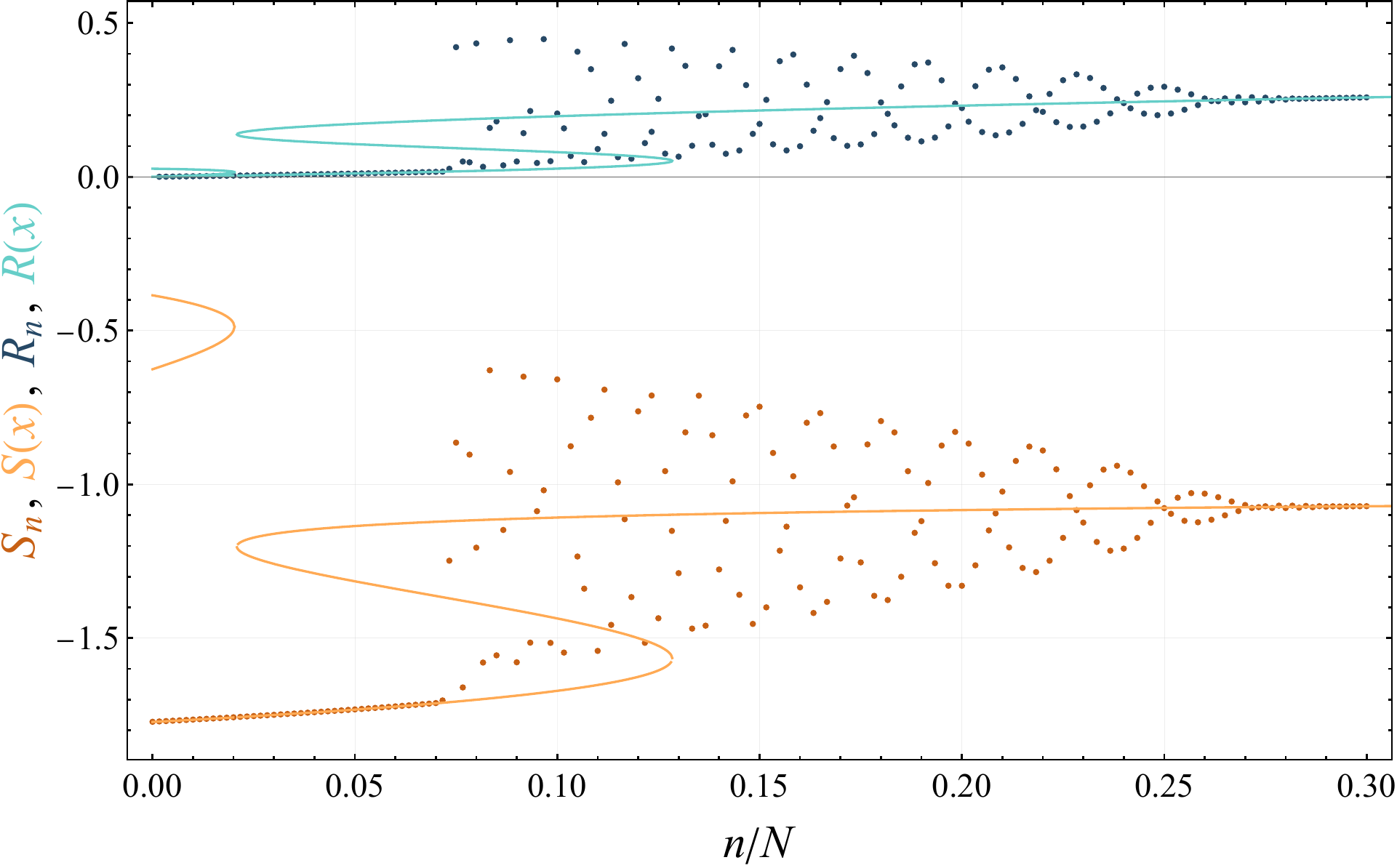}}
\hfill
\subfigure[$w_3=4, \alpha=-1, \beta=0$]{
  \includegraphics[width=0.31\textwidth]{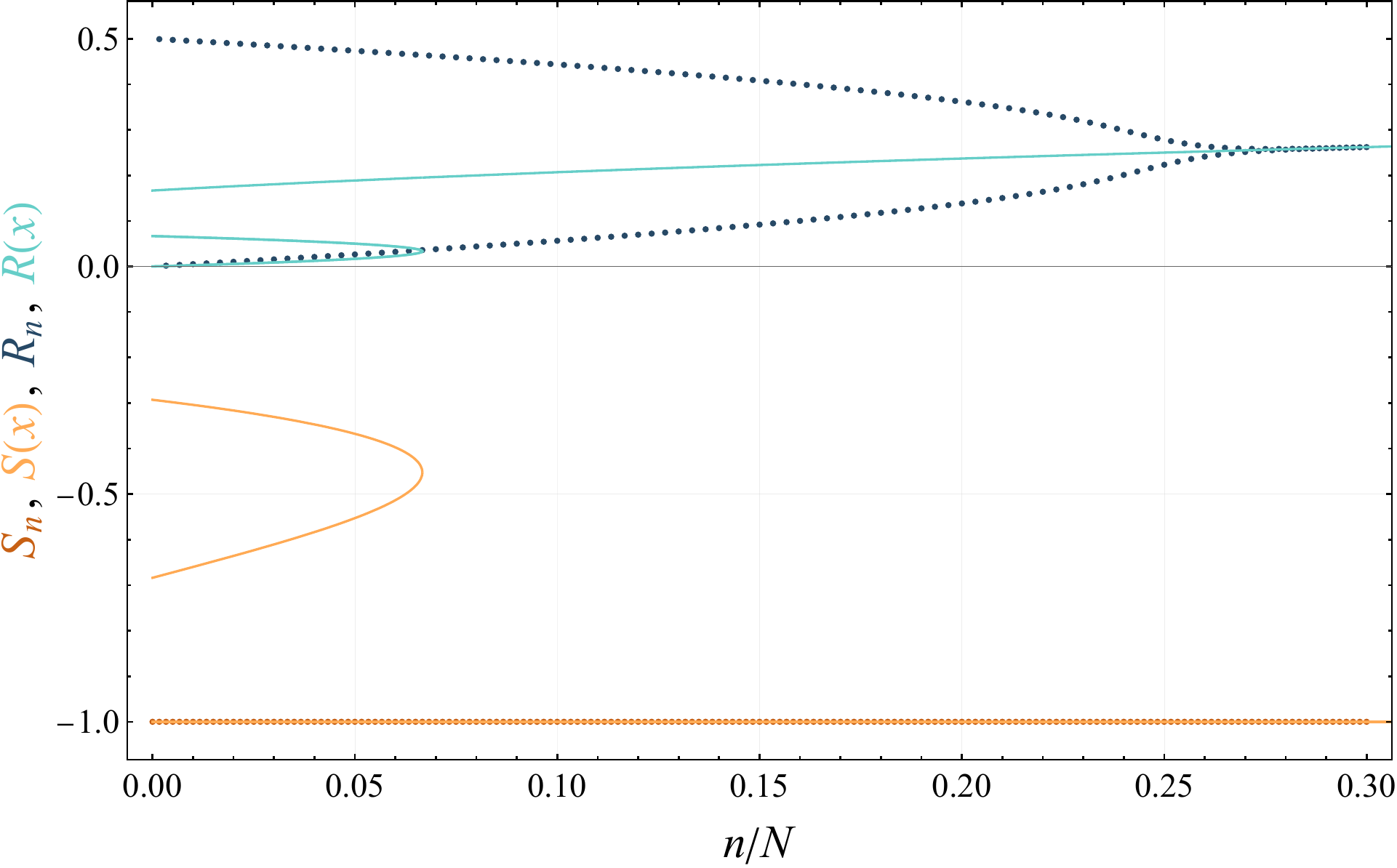}}

\par\medskip

\subfigure[$w_3=4, \alpha=-1, \beta=-\sqrt{-\gamma_1}$]{
  \includegraphics[width=0.31\textwidth]{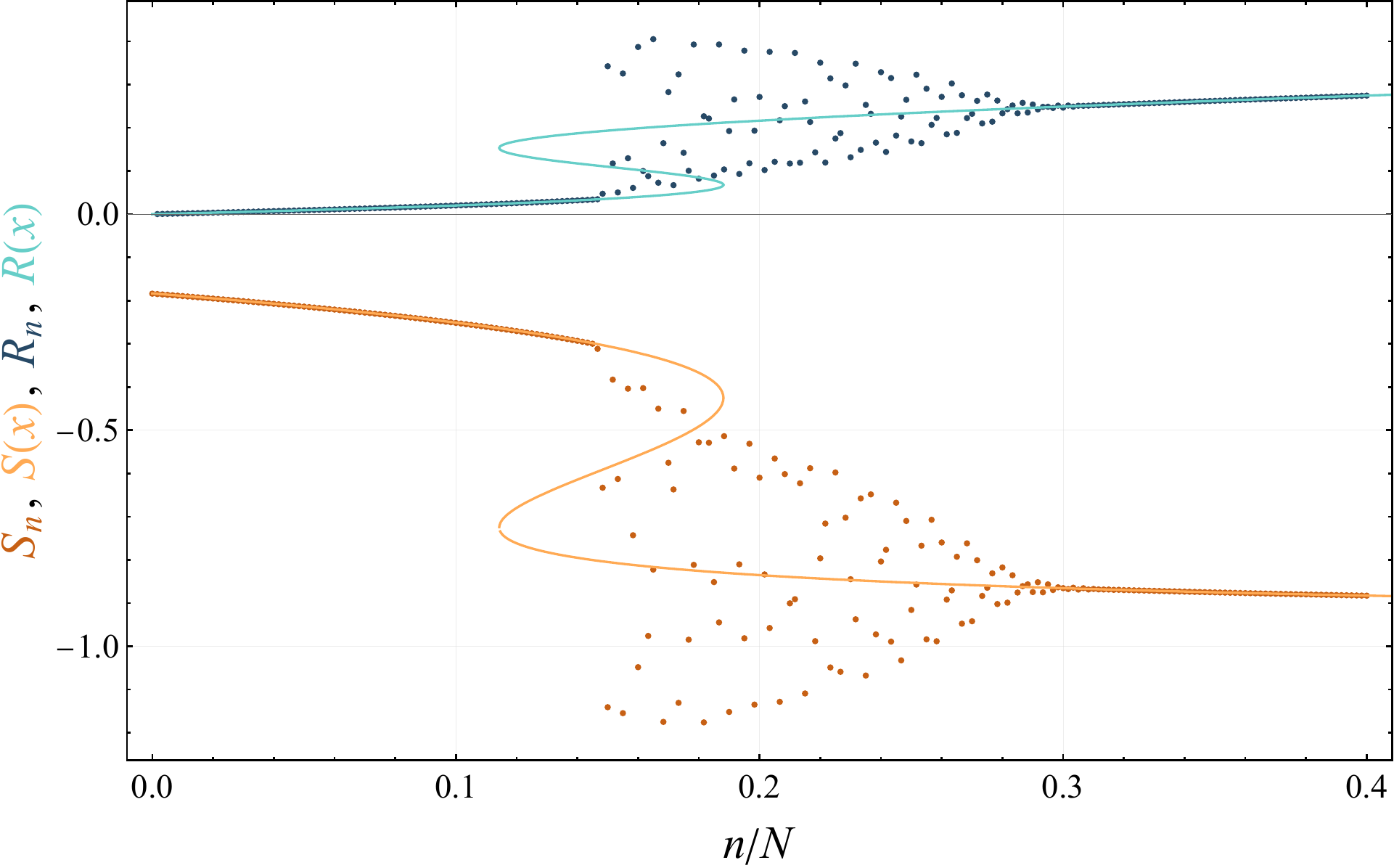}}
\hfill
\subfigure[$w_3=4, \alpha=-1, \beta=-0.3$]{
  \includegraphics[width=0.31\textwidth]{figure/_0.3B.pdf}}
\hfill
\subfigure[$w_3=4, \alpha=1, \beta=0.3$]{
  \includegraphics[width=0.31\textwidth]{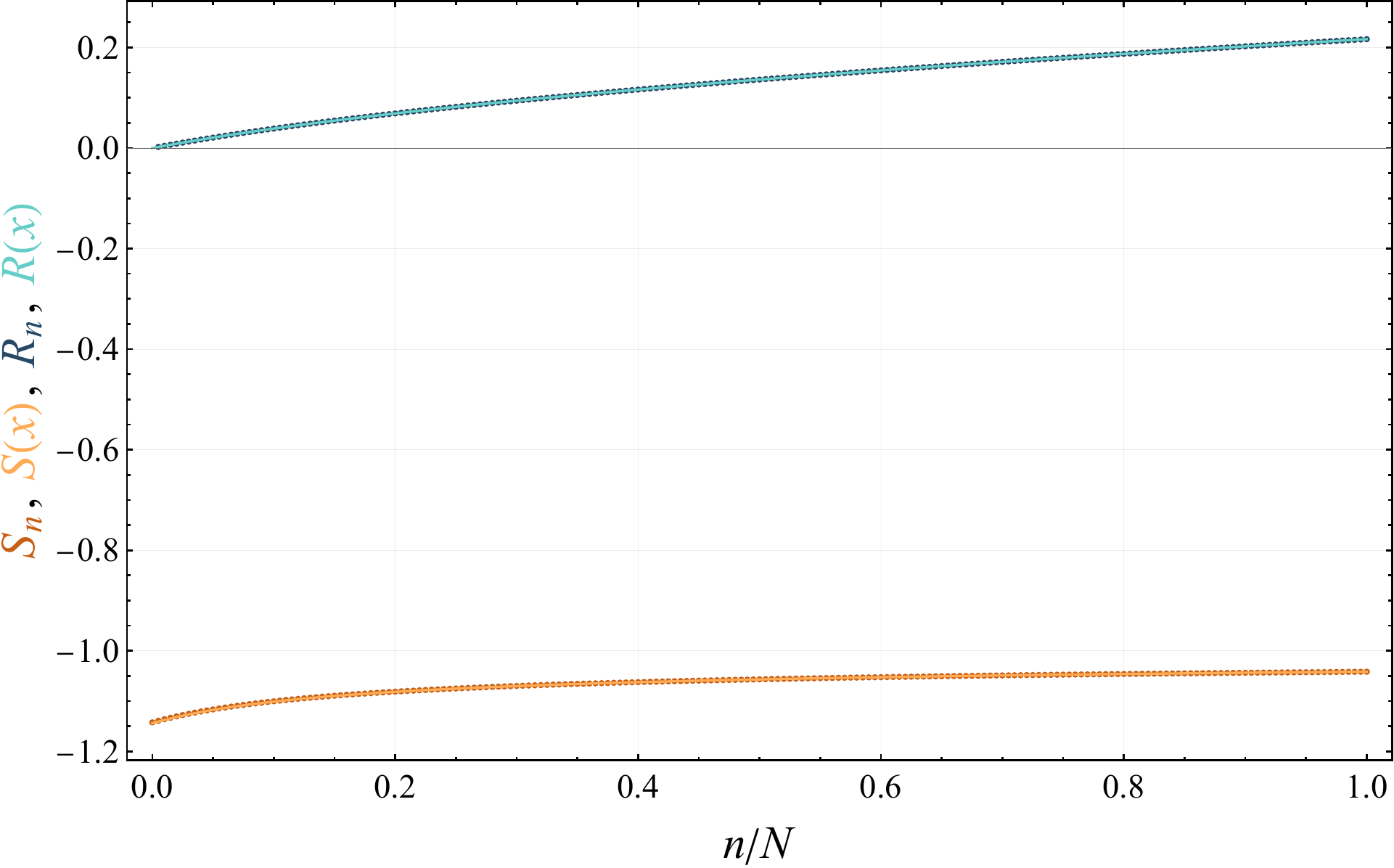}}
\caption{Recursion coefficients $R_n$ and $S_n$ for $w_3=4$, $|\alpha|=1$
and representative values of $\beta$ in the range $-1.5\leq\beta\leq1.5$, compared
with the recursion functions $R(x)$ and $S(x)$.}
\label{recurrencecoefficientquamain}
\end{figure}
In view of the convergence of the recursion coefficients to the recursion functions in the large-$n$ limit, it is instructive to examine the continuum string equations satisfied by recursion functions. For the quartic potential
$V_q$, they take the form
\begin{equation}\label{continuumstringequationsqua}
\begin{aligned}
4\left(S^3+6RS\right)+3w_3\left(S^2+2R\right)+2w_2S+w_1 &= 0,\\
12R^2+12RS^2+6w_3RS+2w_2R &= x.
\end{aligned}
\end{equation}
With the shifted variable $U=S+w_3/4$ and the effective couplings
\begin{equation}
\alpha=w_2-\frac{3w_3^2}{8},
\qquad
\beta=w_1-\frac{w_2w_3}{2}+\frac{w_3^3}{8},
\label{eq:effective-couplings}
\end{equation}
the continuum string equations reduce to
\begin{align}
4U^3+24RU+2\alpha U+\beta&=0,
\label{eq:reduced-string-S}\\
12R^2+12RU^2+2\alpha R&=x.
\label{eq:reduced-string-R}
\end{align}
Eliminating $R$ from the equations above gives
\begin{equation}
x
=
-
\frac{
\left(20U^3+2\alpha U-\beta\right)
\left(4U^3+2\alpha U+\beta\right)
}
{48U^2}.
\label{eq:x-parametric}
\end{equation}
At the turning point of $x(U)$, the derivatives of $R(x)$ and $S(x)$ diverge, the
phenomenon known as the ``gradient catastrophe''~\cite{brezin1990non,clarkson2025symmetric}.
Such points are therefore determined by
\begin{equation}
\frac{dx}{dU}
=
-
\frac{
160U^6
+
48\alpha U^4
+
8\beta U^3
+
\beta^2
}
{24U^3}
=
0.
\end{equation}
As shown below, these singular points have a pronounced effect on the recursion
coefficients.  
\begin{figure}[!htbp]
\centering

\subfigure[$w_3=4, \alpha=-1, \beta=1.5$]{
  \includegraphics[width=0.31\textwidth]{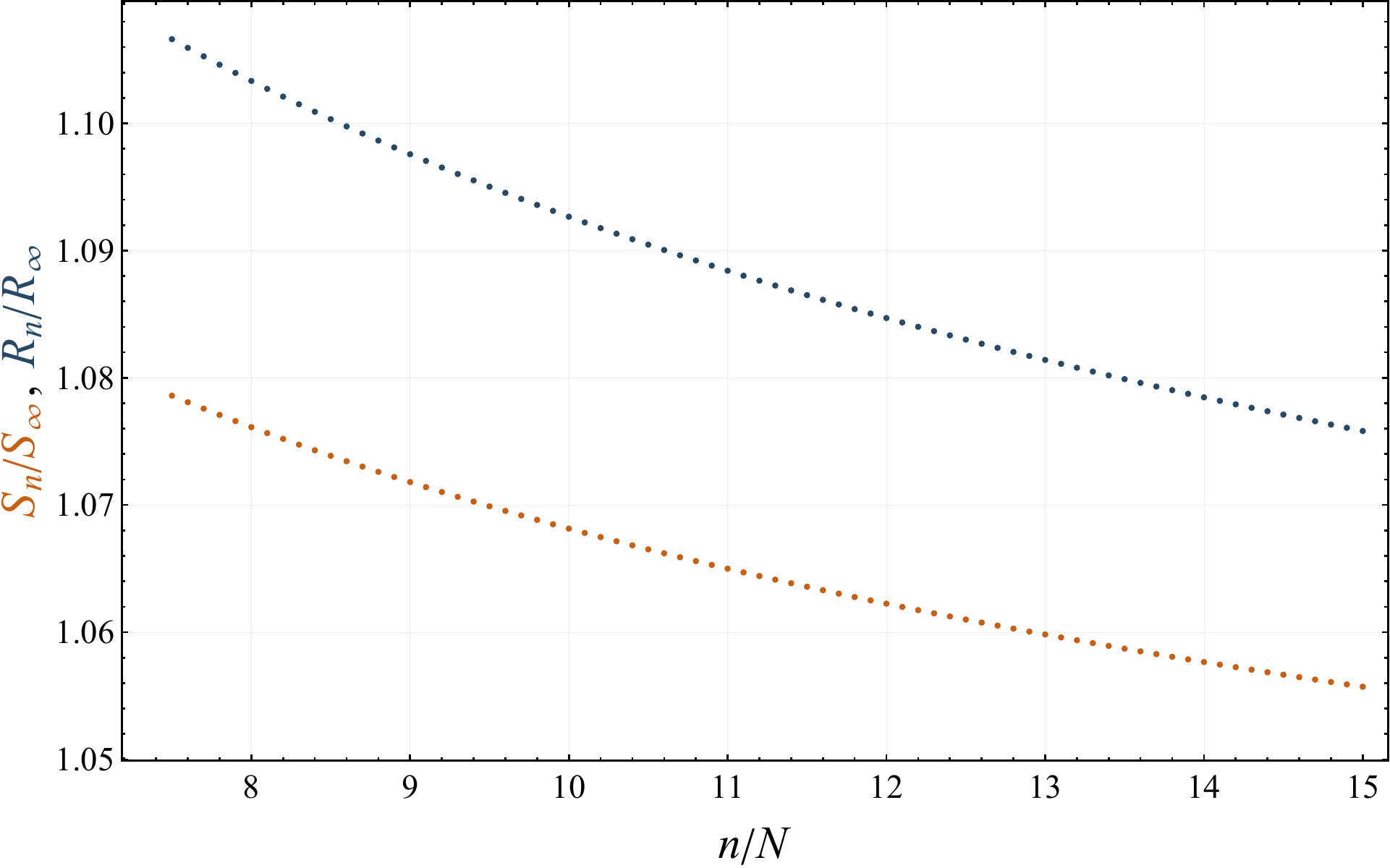}}
\hfill
\subfigure[$w_3=4, \alpha=-1, \beta=1$]{
  \includegraphics[width=0.31\textwidth]{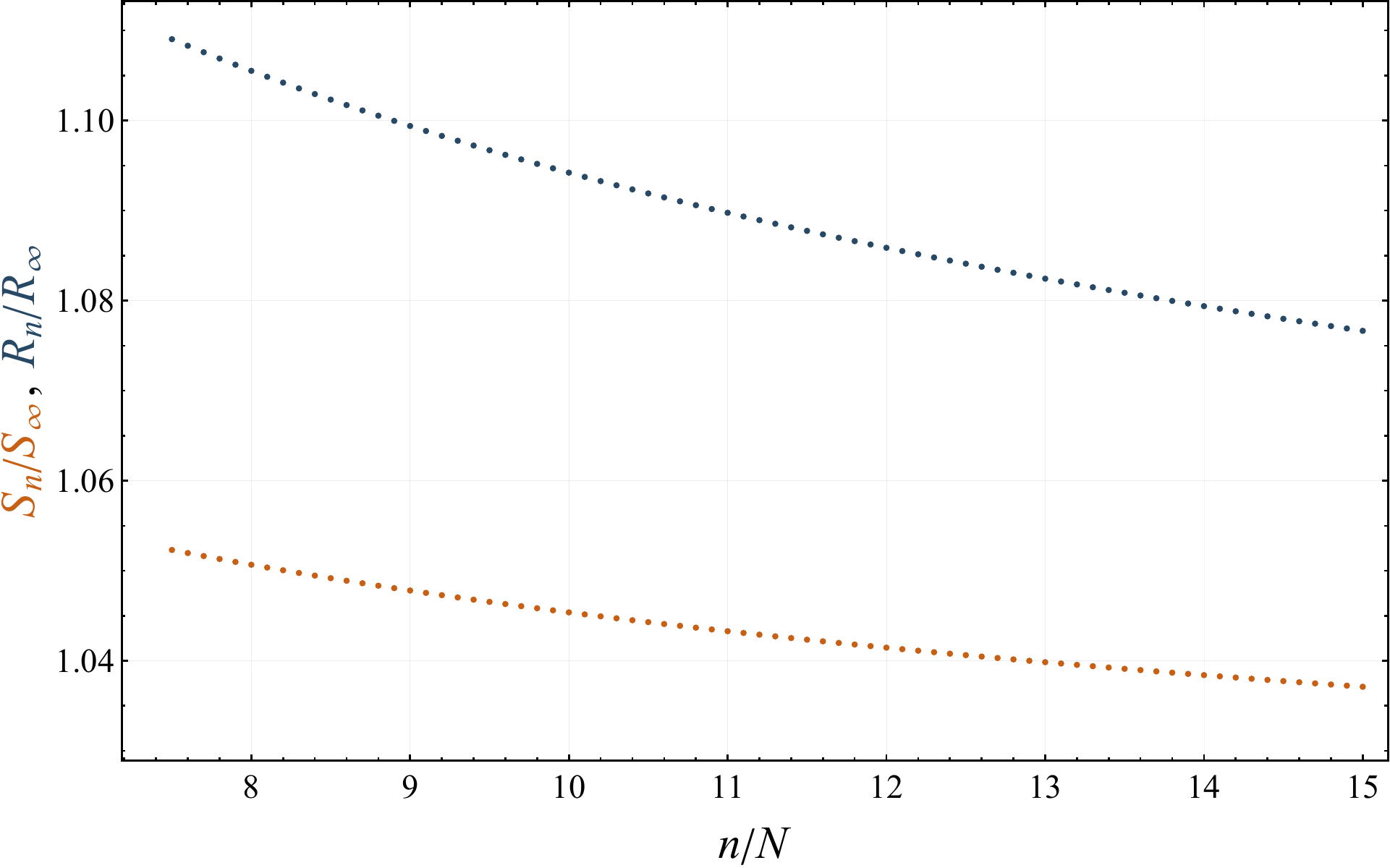}}
\hfill
\subfigure[$w_3=4, \alpha=-1, \beta=0.3$]{
  \includegraphics[width=0.31\textwidth]{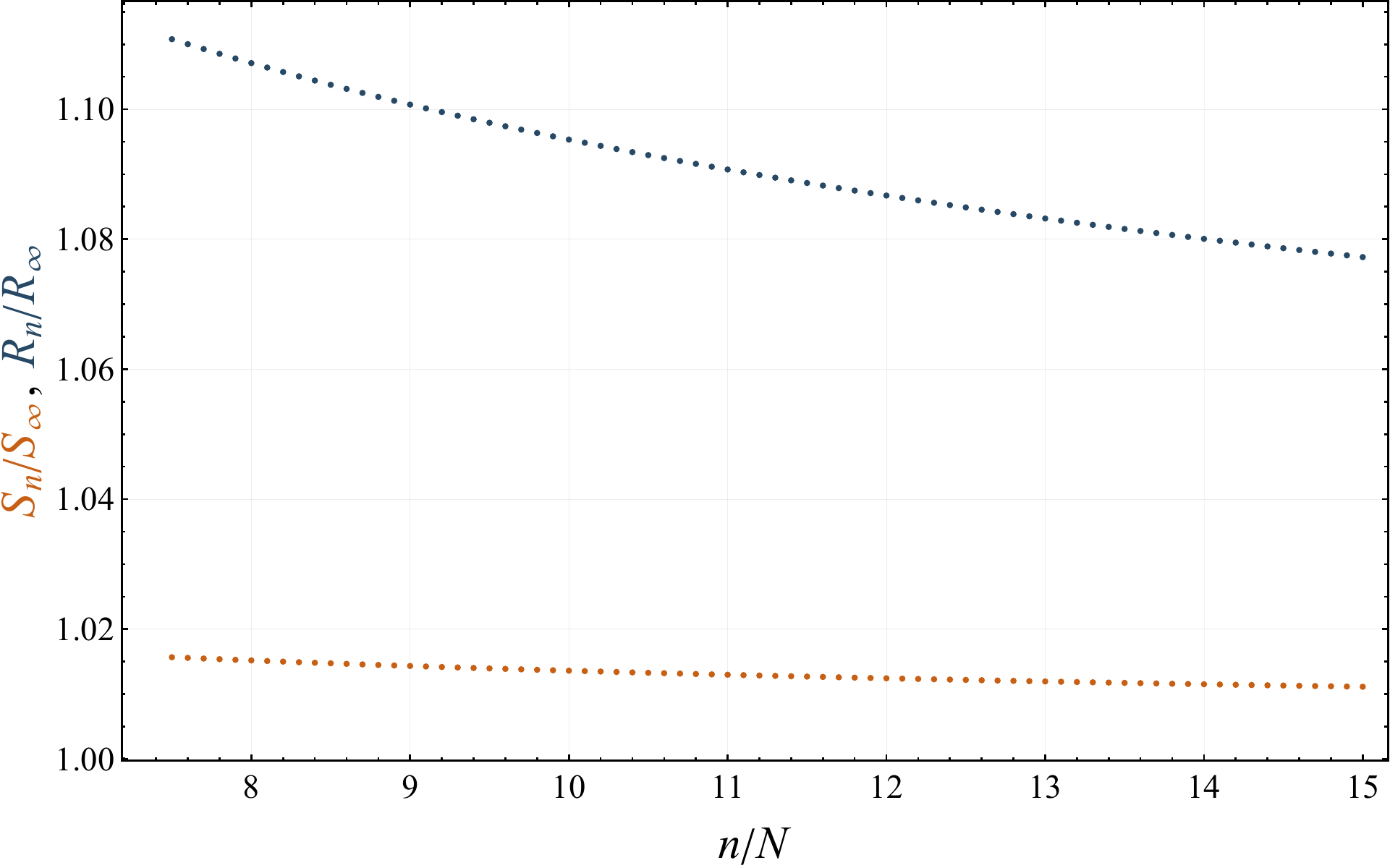}}
\caption{Numerical verification of the asymptotic behavior predicted by
Eq.~\eqref{eq:lanczos-asymptoticsqu} for $w_3=4$ and $\alpha=-1$.  The ratios
$R_n/R_\infty$ and $S_n/S_\infty$ are shown for representative values of
$\beta$.}
\label{fig:quartic-ratio}
\end{figure}
The number of such points, equivalently the number of real roots
of the equation above, partitions the $(\alpha,\beta)$ parameter space into
regions with distinct root structures.  The discriminant of the numerator
polynomial is proportional to
\begin{equation}
\Delta
\propto
\beta^6
\left(8\alpha^3+27\beta^2\right)
\left(256\alpha^3+135\beta^2\right).
\end{equation}
The boundaries between these regions occur at the zeros of this discriminant.
In terms of the ratio $\gamma=\beta^2/\alpha^3$, they correspond to the critical
values
\begin{equation}
\gamma_0=0,
\qquad
\gamma_1=-\frac{8}{27},
\qquad
\gamma_2=-\frac{256}{135}.
\label{eq:h1-h2}
\end{equation}
Before presenting the numerical results, we note that the recursion functions
$R(x)$ and $S(x)$ depend only on the scaling variable $x=n/N$ and are therefore
invariant under the simultaneous rescaling
\begin{equation}
N\to \epsilon N,
\qquad
n\to \epsilon n.
\end{equation}
Although this scaling invariance does not hold exactly for the recursion coefficients, Fig.~\ref{invariantundertransition} shows that their
qualitative behavior remains stable as $N$ increases. 
\begin{figure}[!htbp]
\centering

\subfigure[$w_3=4, \alpha=-1, \beta=0.3$]{
  \includegraphics[width=0.45\textwidth]{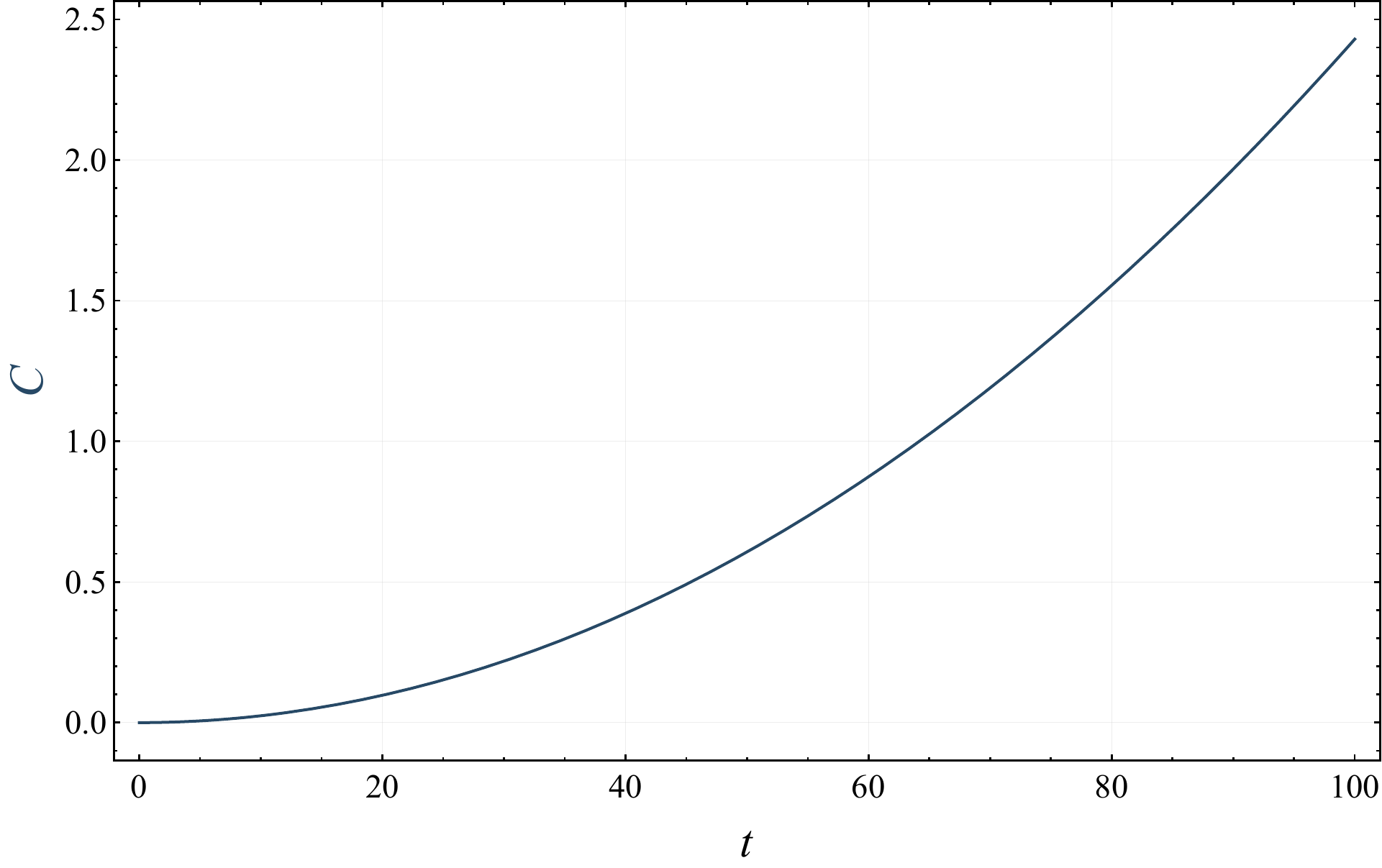}}
\hfill
\subfigure[$w_3=4, \alpha=-1, \beta=0$]{
  \includegraphics[width=0.45\textwidth]{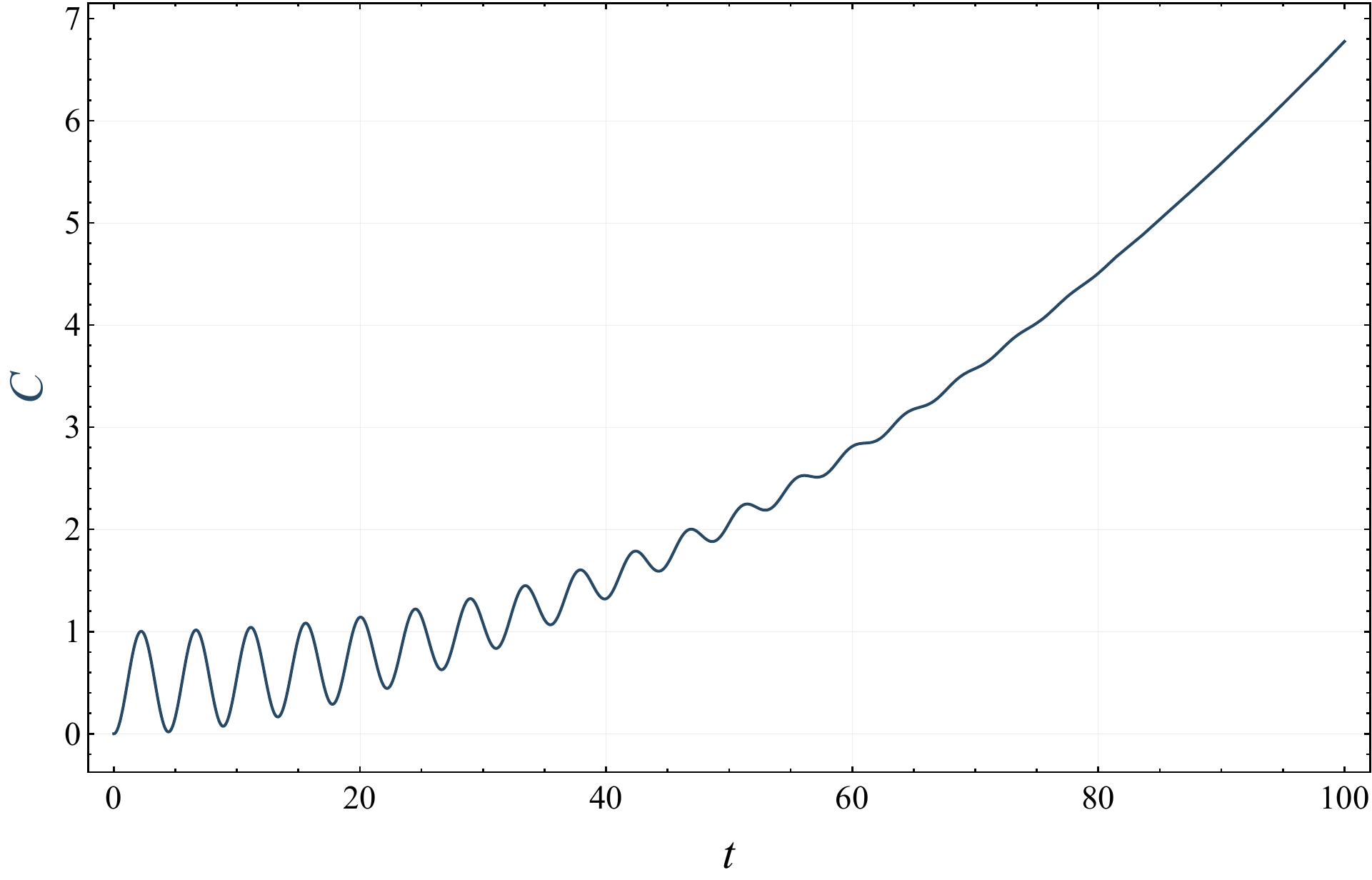}}
\caption{Spread complexity for $w_3=4$ and $\alpha=-1$ at $N=800$.  For
$\beta=0.3$, the recursion coefficients contain a transition region, as shown
in Fig.~\ref{recurrencecoefficientquamain}, while the corresponding complexity
remains smooth and increases monotonically, as in the generic case.  For
$\beta=0$, the splitting of the recursion coefficients into two branches leads
to early time oscillations followed by monotonic growth.}
\label{fig:quartic-krylov-complexity}
\end{figure}
Throughout the remainder of this work, the recursion coefficients are plotted
against $n/N$, with $N$ specified only when necessary. Figure~\ref{recurrencecoefficientquamain} compares the recursion coefficients
$R_n$ and $S_n$ with the recursion functions $R(x)$ and $S(x)$ for $w_3=4$,
$|\alpha|=1$ and $-1.5\leq\beta\leq1.5$.  As expected, the recursion functions
reproduce the large-$n/N$ behavior of the discrete coefficients.  Remarkably,
they also provide a good approximation at small $n/N$.  As in the symmetric
case, $R_n$ develops a ``chaotic'' transition region after departing from
$R(x)$.  In the asymmetric case, $S_n$ can depart from $S(x)$ and
enter a ``chaotic'' transition region simultaneously with $R_n$. As $\beta$ is varied so that $\gamma=\beta^2/\alpha^3$ crosses the critical values
in Eq.~\eqref{eq:h1-h2}, these transition regions emerge, whereas at $\beta=0$ they
disappear and the recursion coefficients split into two branches.  Each
transition region contains at least one gradient catastrophe, which motivates
classifying the parameter space according to the number of such singularities.
Conversely, in parameter regimes where no gradient catastrophe occurs, the
recursion coefficients remain well described by the recursion functions.  In
all cases shown, $R_n$ increases monotonically at large $n/N$, whereas $S_n$ increases for positive $\beta$ and decreases for negative $\beta$. The asymptotic behavior derived in section~\ref{asymptotics} can be tested
directly for the quartic potential $V_q$.  In this case,
Eq.~\eqref{eq:lanczos-asymptotics} reduces to
\begin{equation}\label{eq:lanczos-asymptoticsqu}
R_\infty(n)=\sqrt{\frac{n}{12N}},
\quad
S_\infty=-\frac{w_3}{4}.
\end{equation}
Figure~\ref{fig:quartic-ratio} confirms that both $R_n/R_\infty$ and
$S_n/S_\infty$ approach one as $n/N\to\infty$. Finally, we compute the spread
complexity using the Schr\"odinger equation~\eqref{psieq}.  Since the Krylov
amplitudes do not possess the scaling
invariance discussed above, the value of $N$ must be specified explicitly.
The transition region in the recursion coefficients does not qualitatively
alter the spread complexity.  For $\beta=0.3$, the complexity remains smooth
and increases monotonically.  By contrast, at $\beta=0$ the two branch
structure of the recursion coefficients produces early time oscillations
before the complexity crosses over to monotonic growth.

\section{Double-scaled SYK model}
\label{sec:effective-potential}
The Sachdev--Ye--Kitaev (SYK) model describes $M$ Majorana fermions with
all to all random $p$ body interactions~\cite{Maldacena:2016hyu}.  Its
Hamiltonian is
\begin{equation}
H = i^{p/2} \sum_{1 \leq i_1 < \cdots < i_p \leq M}
J_{i_1 \cdots i_p}\,\psi_{i_1}\cdots\psi_{i_p},
\end{equation}
where $p$ is even and the couplings $J_{i_1\cdots i_p}$ are Gaussian random
variables with vanishing mean and fixed variance.  The operators $\psi_i$
satisfy the Majorana anticommutation relations.  The double-scaled limit is
obtained by taking $p,M\to\infty$ while keeping
$q=e^{-2p^2/M}\in(0,1)$
fixed~\cite{Cotler:2016fpe,Garcia-Garcia:2018fns,Berkooz:2018jqr}.  A matrix
potential associated with the density of states of the double-scaled SYK
(DSSYK) model admits the exact Chebyshev expansion~\cite{Jafferis:2022wez}
\begin{equation}
V_s(\lambda)
=
\sum_{n=1}^{\infty}
\frac{(-1)^{n-1}}{n}
q^{n^2/2}
\left(
q^{n/2}+q^{-n/2}
\right)
T_{2n}
\!\left(
\frac{\sqrt{1-q}}{2}\lambda
\right),
\label{eq:q-matrix-potential}
\end{equation}
where $T_n$ denotes the Chebyshev polynomial of the first kind.  
\begin{figure}[!htbp]
\centering

\subfigure[$q=1-10^{-5}$]{
  \includegraphics[width=0.31\textwidth]{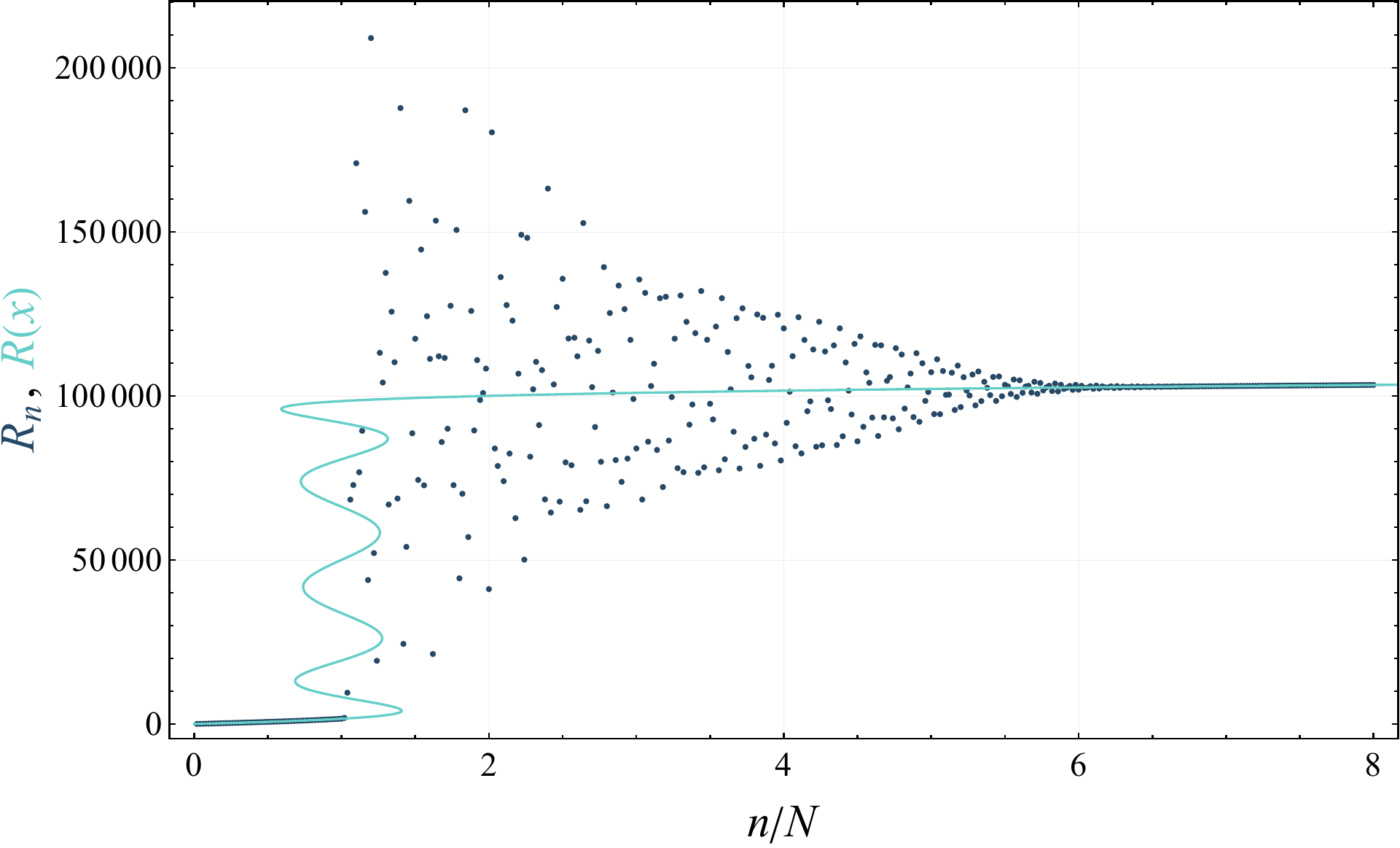}}
\hfill
\subfigure[$q=0.999$]{
  \includegraphics[width=0.31\textwidth]{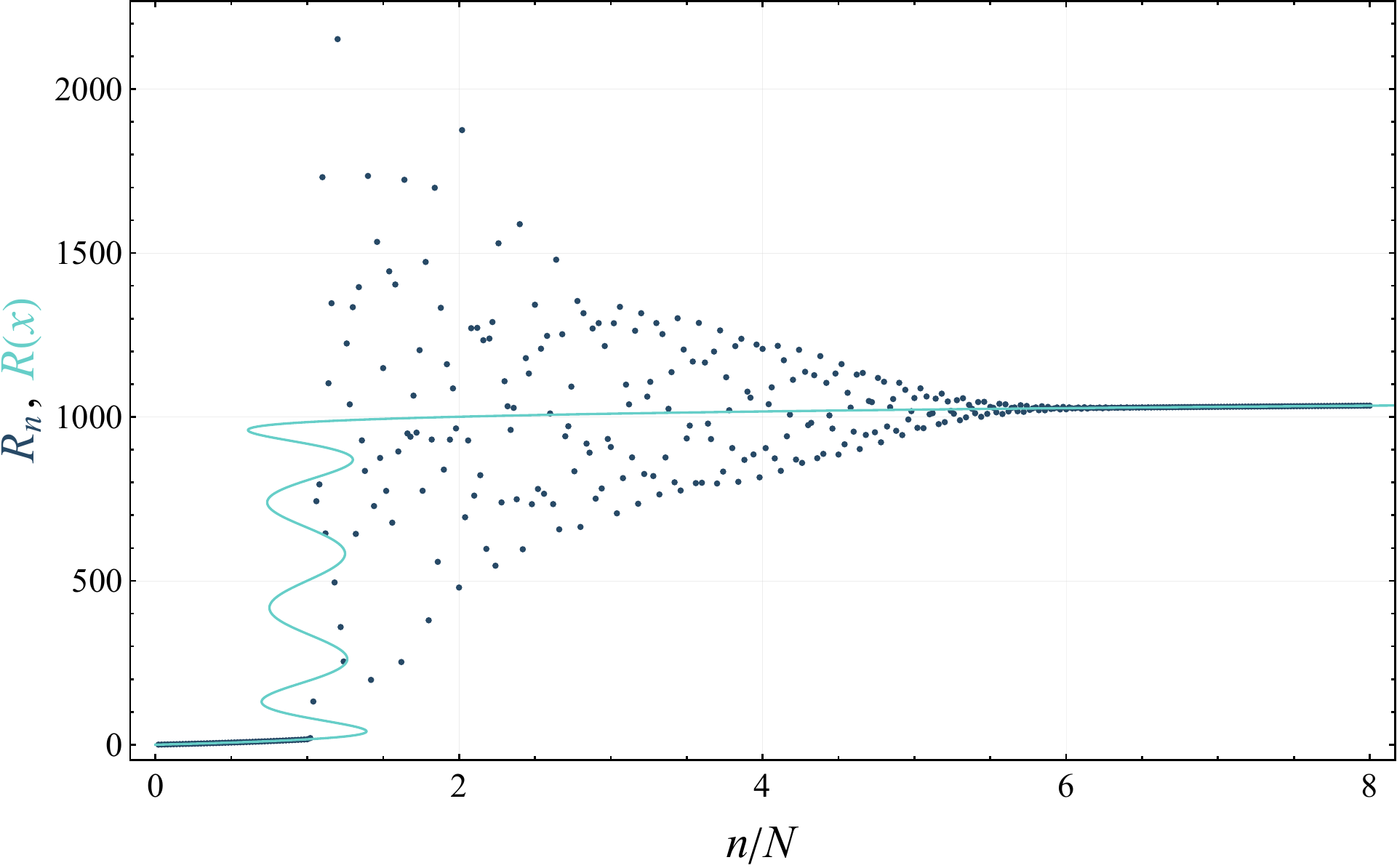}}
\hfill
\subfigure[$q=0.99$]{
  \includegraphics[width=0.31\textwidth]{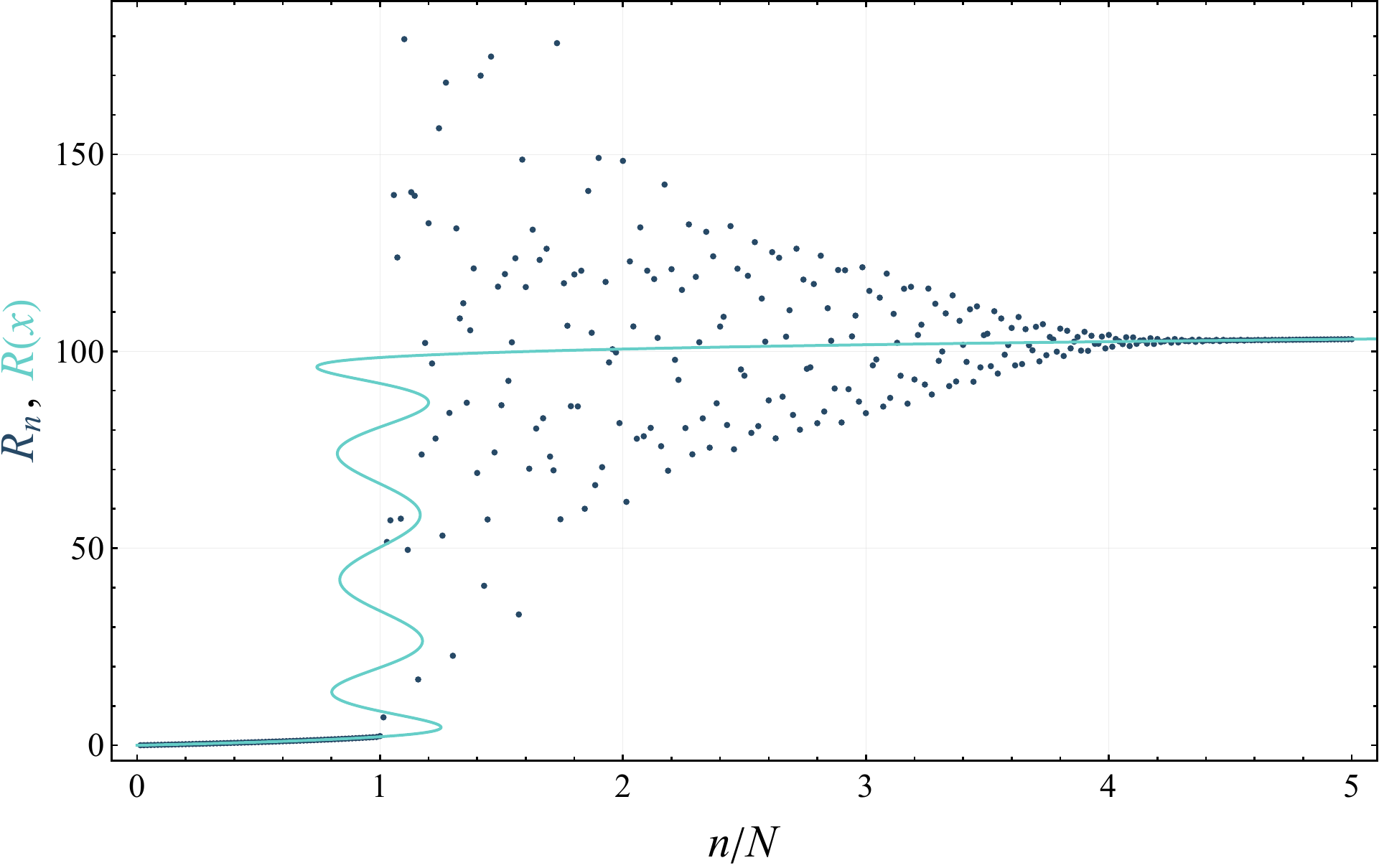}}

\par\medskip

\subfigure[$q=0.94$]{
  \includegraphics[width=0.31\textwidth]{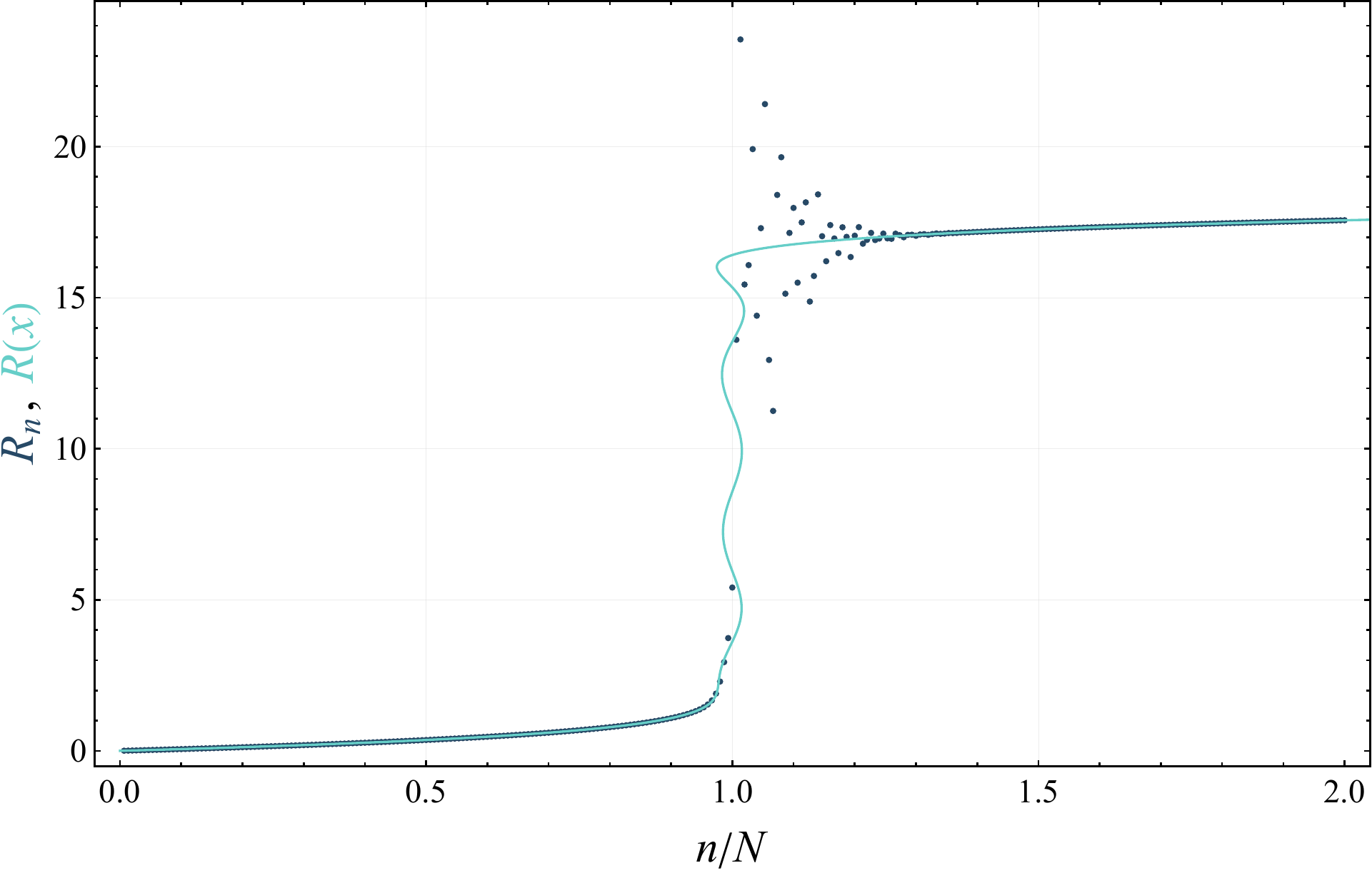}}
\hfill
\subfigure[$q=0.88$]{
  \includegraphics[width=0.31\textwidth]{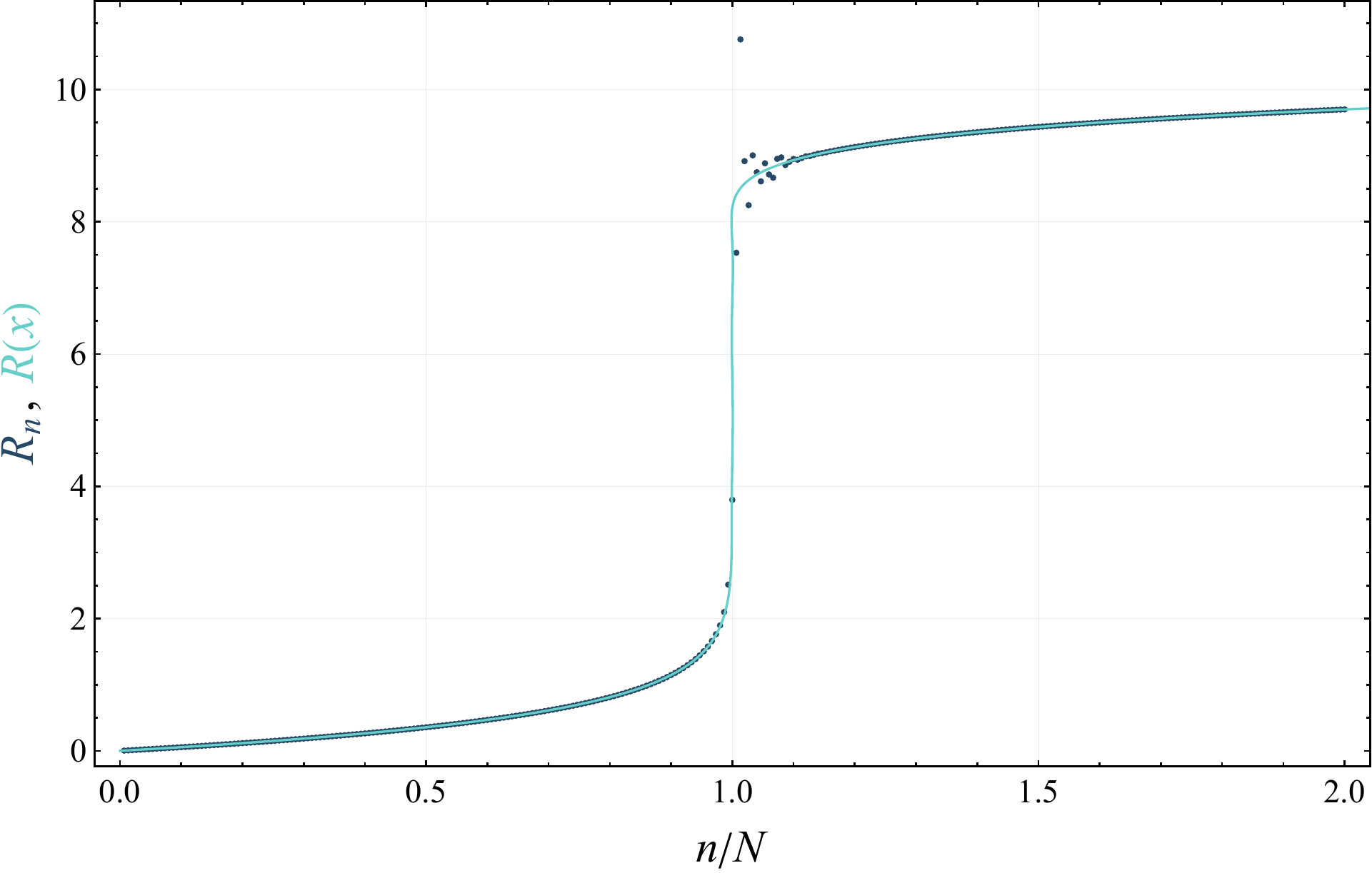}}
\hfill
\subfigure[$q=0.81$]{
  \includegraphics[width=0.31\textwidth]{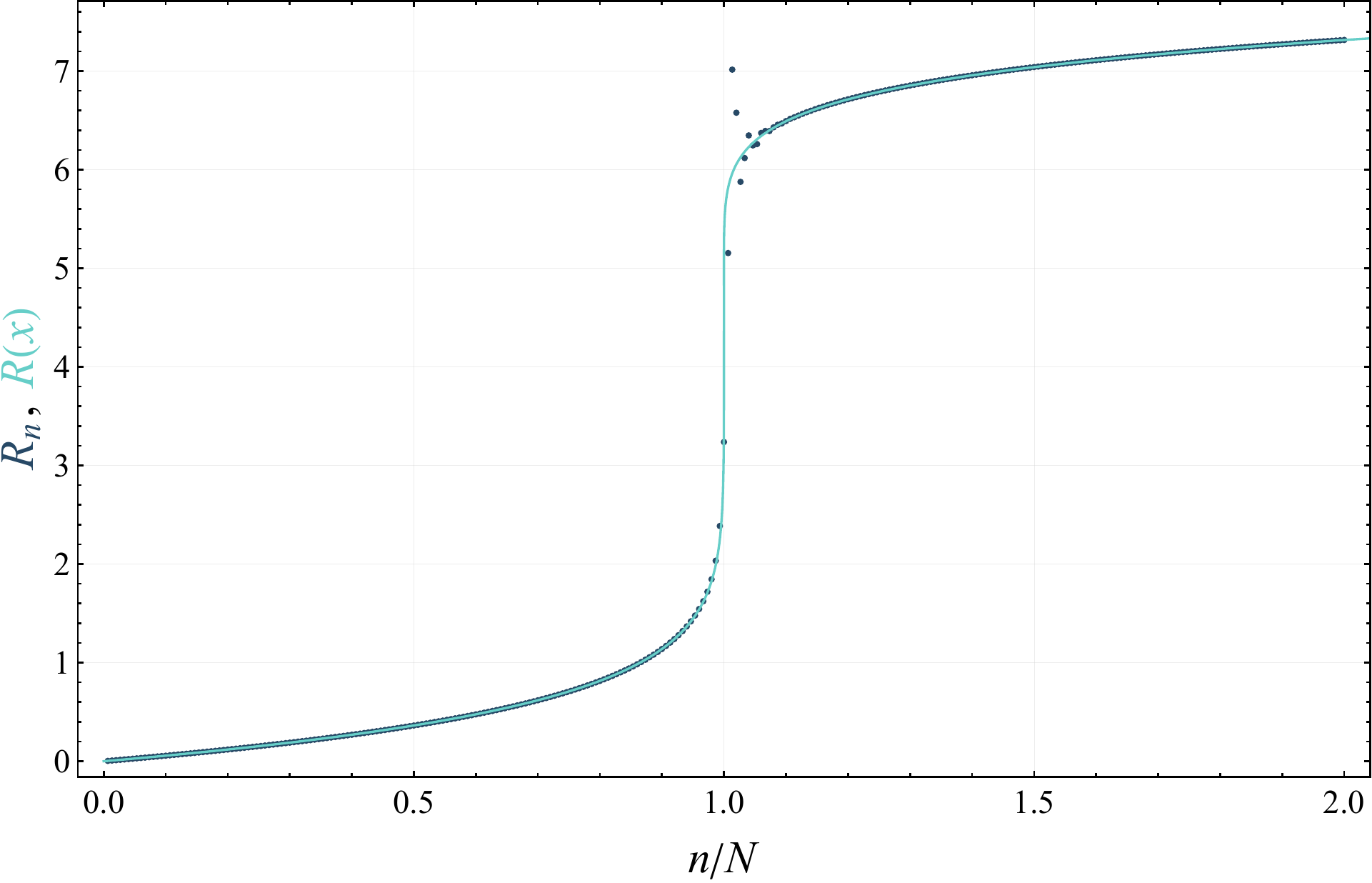}}

\par\medskip

\subfigure[$q=0.75$]{
  \includegraphics[width=0.31\textwidth]{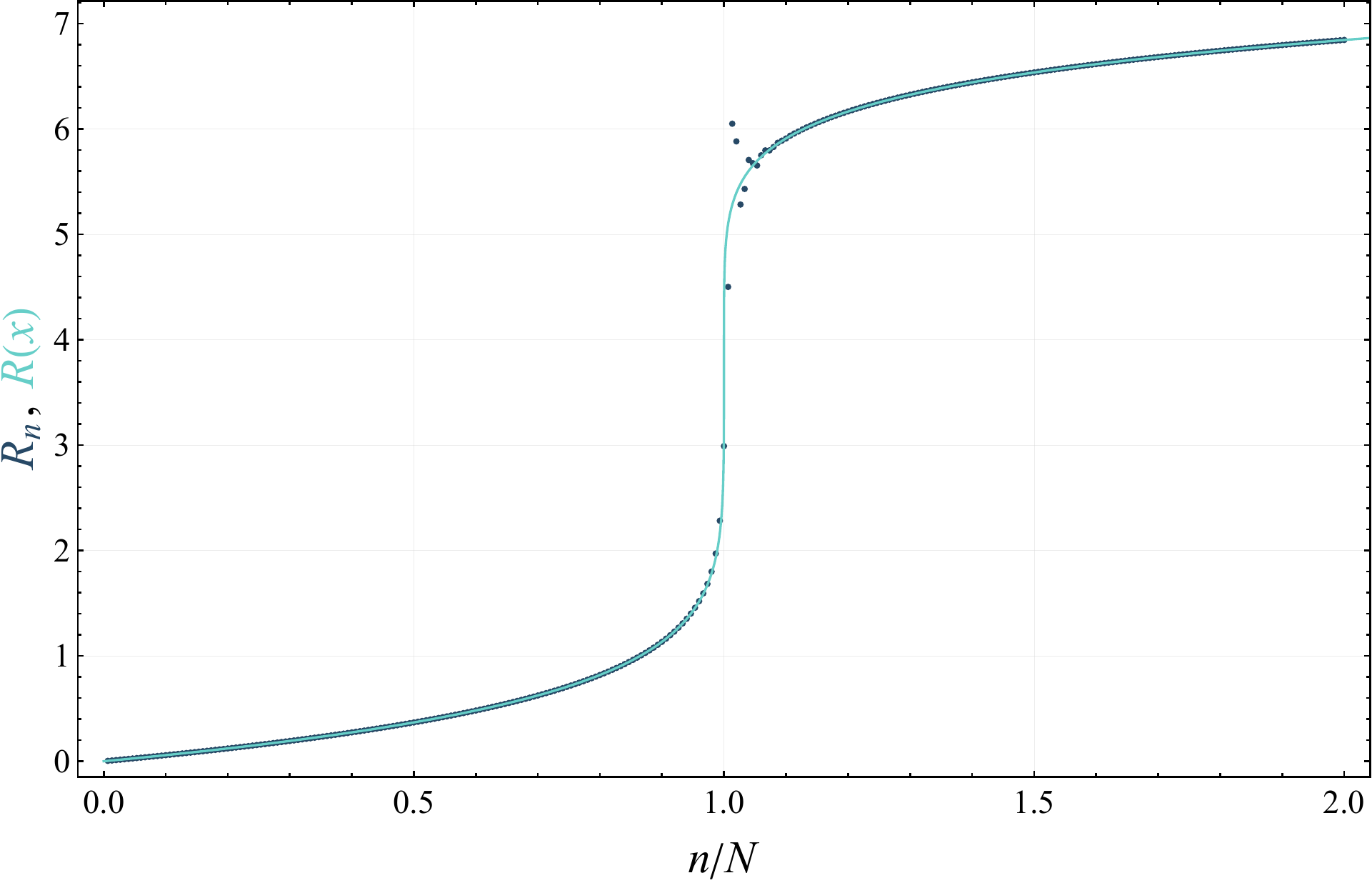}}
\hfill
\subfigure[$q=0.69$]{
  \includegraphics[width=0.31\textwidth]{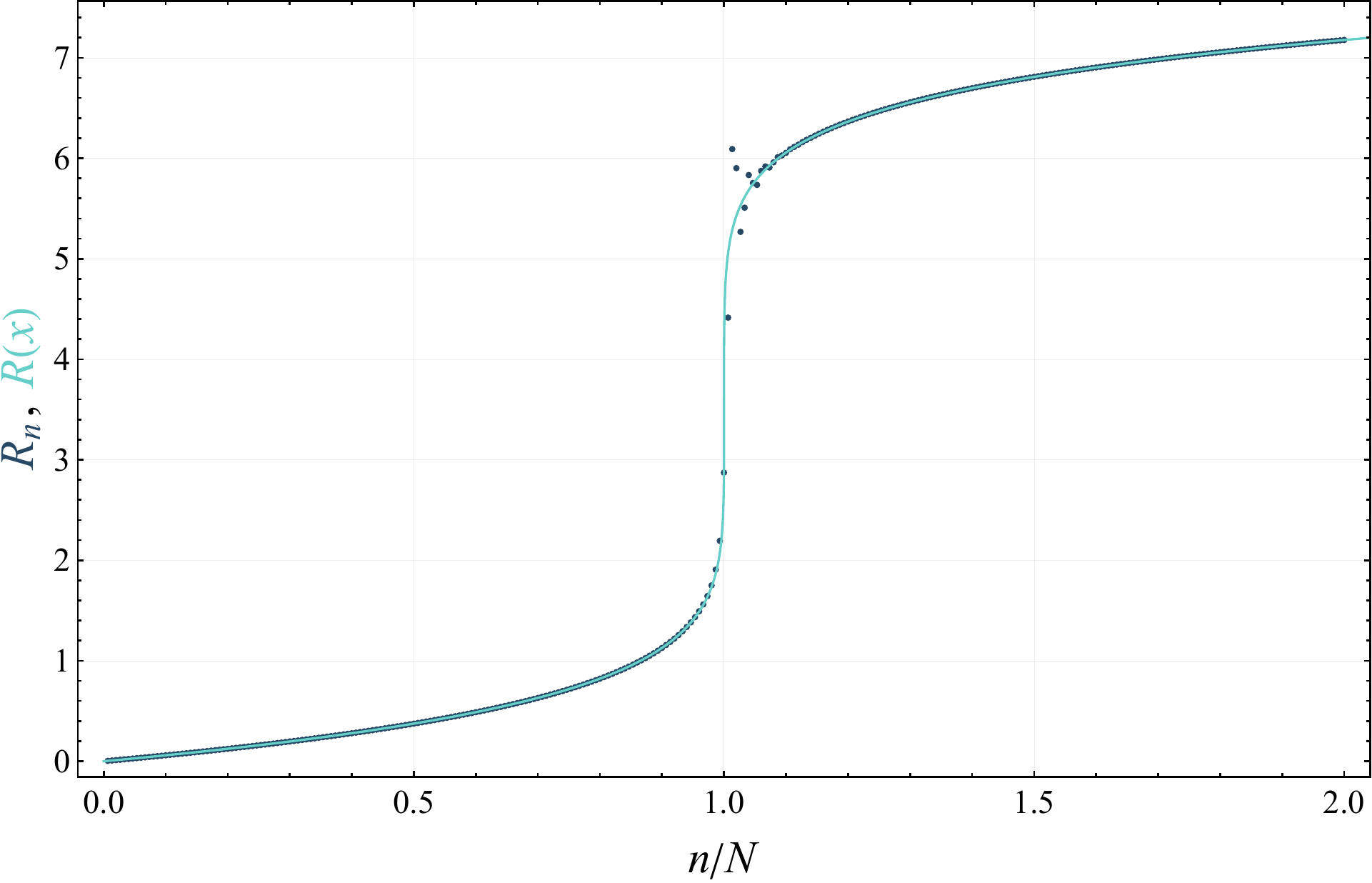}}
\hfill
\subfigure[$q=0.63$]{
  \includegraphics[width=0.31\textwidth]{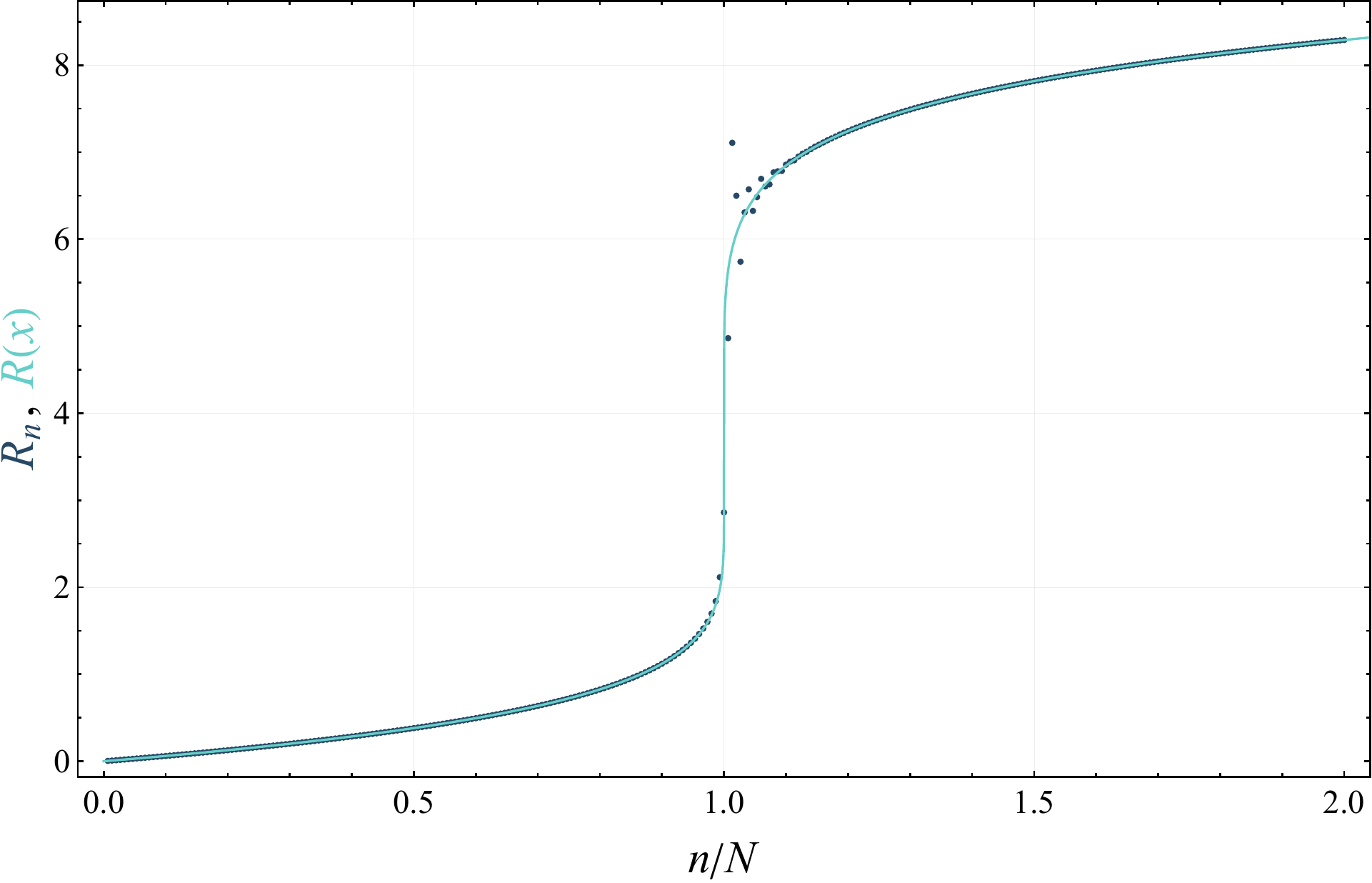}}

\par\medskip

\subfigure[$q=0.52$]{
  \includegraphics[width=0.31\textwidth]{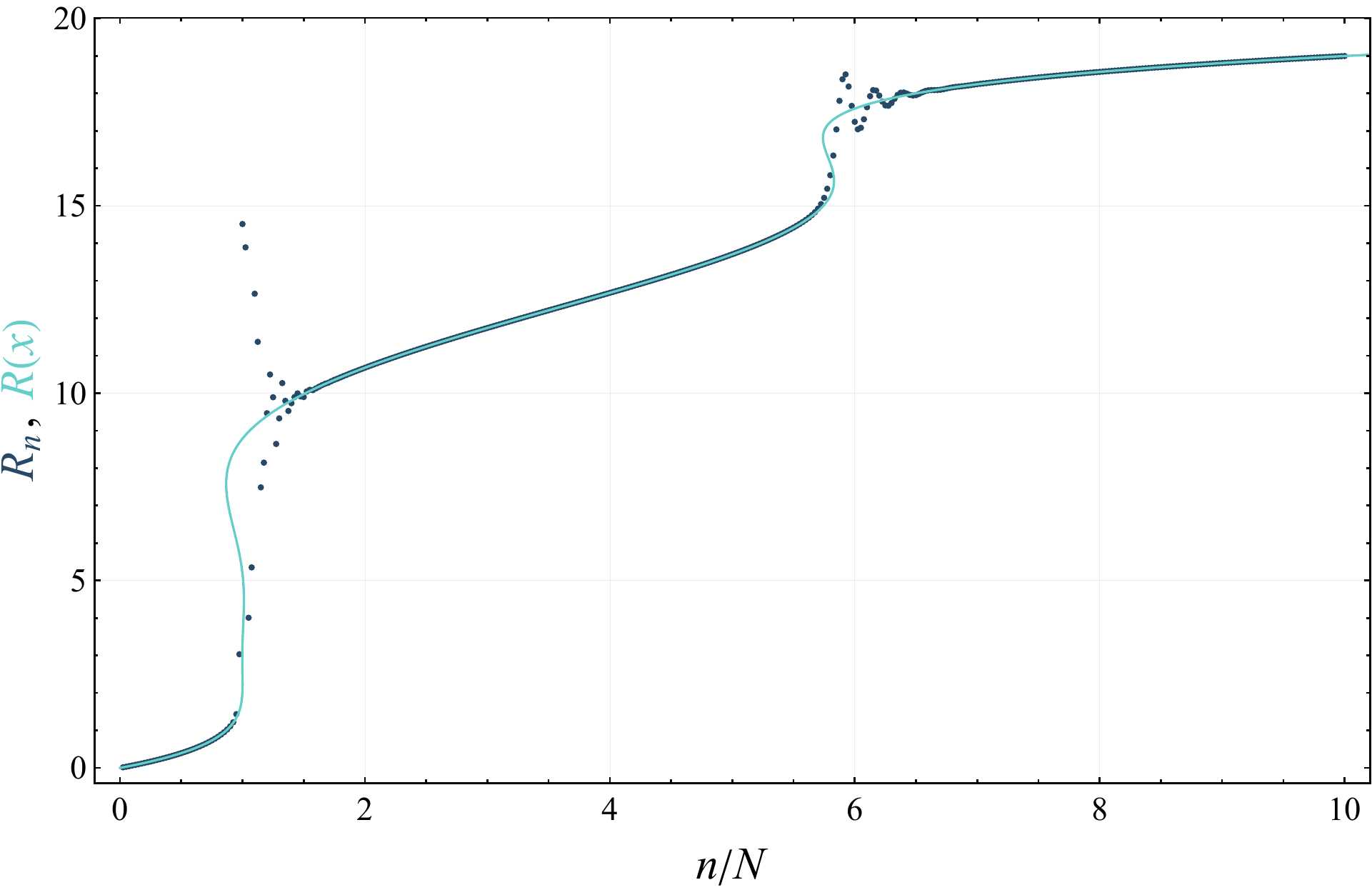}}
\hfill
\subfigure[$q=0.31$]{
  \includegraphics[width=0.31\textwidth]{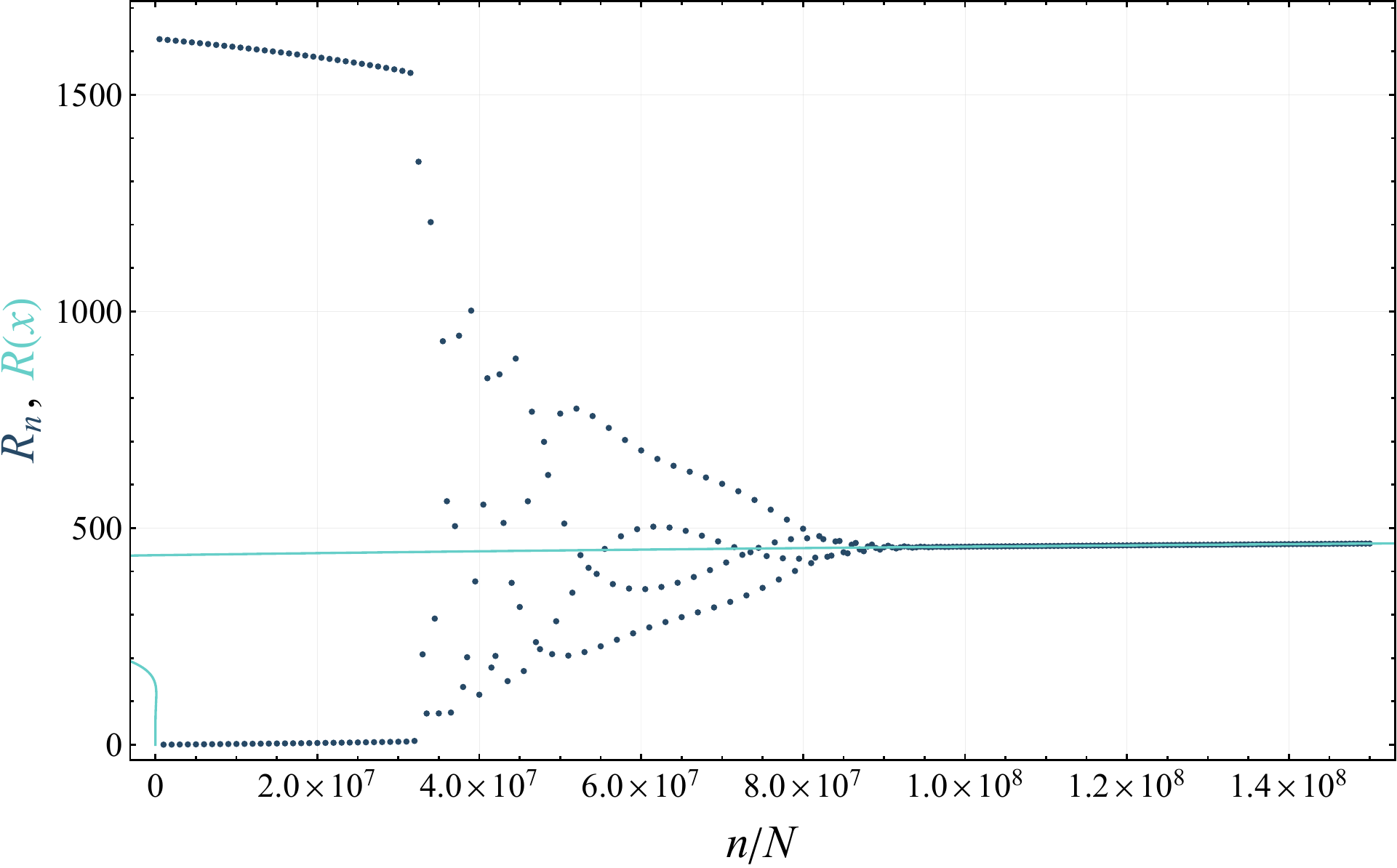}}
\hfill
\subfigure[$q=0.31$]{
  \includegraphics[width=0.31\textwidth]{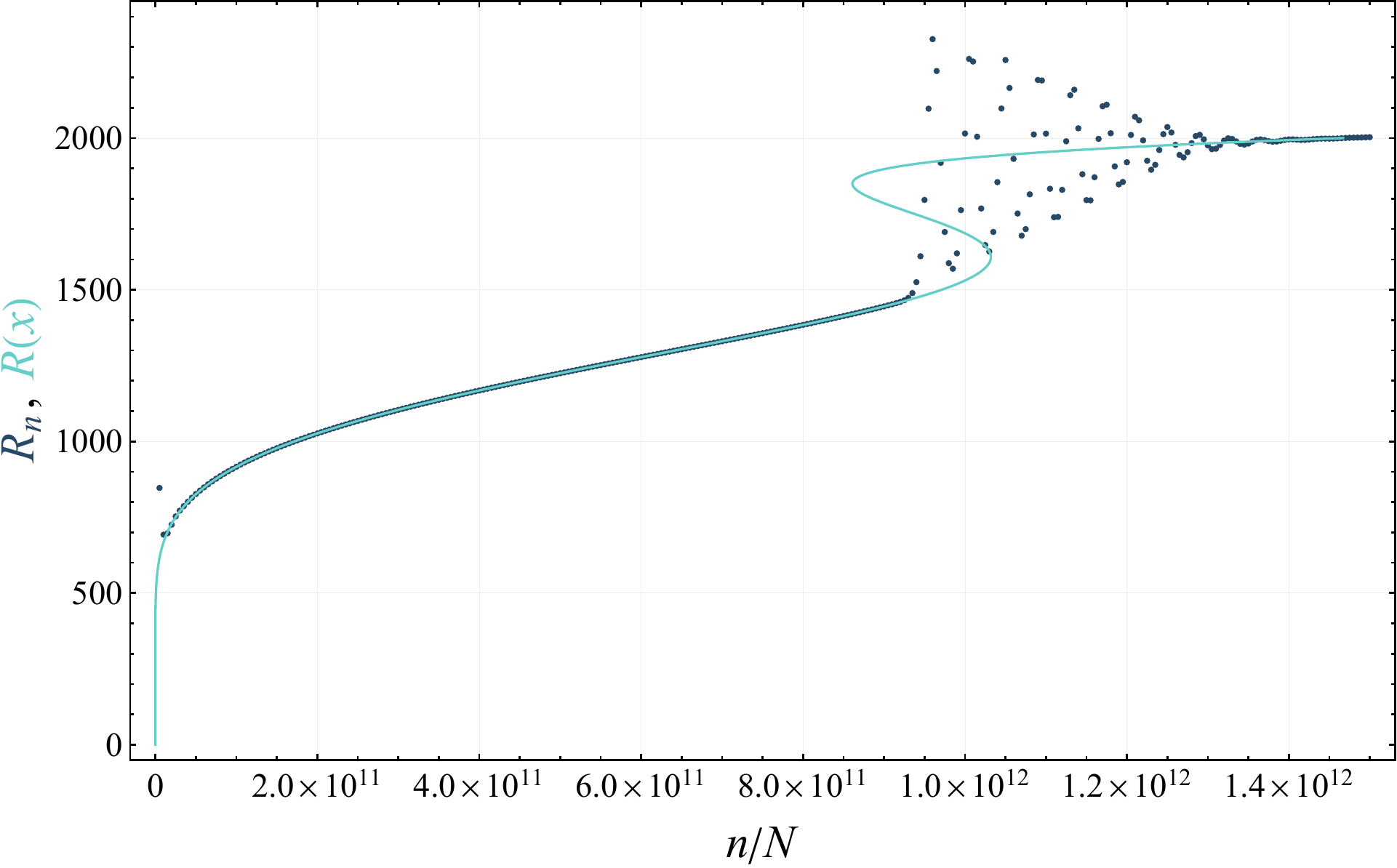}}
\caption{Discrete recursion coefficients $R_n$ for the truncation order $d=18$
and representative values of $q$ in the range $0<q<1$, compared with the
recursion function $R(x)$.}
\label{fig:dssyk-folds}
\end{figure}
For numerical
purposes, the infinite series can be approximated by truncating it at a finite
even degree $d=4k+2$ \footnote{In fact, the recurrence coefficients are highly sensitive to the truncation order. In particular, their asymptotic behavior is entirely determined by the highest retained term, as shown in Eq.~\eqref{eq:lanczos-asymptotics}.} ~\cite{Balasubramanian:2024lqk,Nandy:2024zcd},
\begin{equation}
\label{potentialdssy}
V_s(\lambda)
\approx
\sum_{n=1}^{d/2}
\frac{(-1)^{n-1}}{n}
q^{n^2/2}
\left(
q^{n/2}+q^{-n/2}
\right)
T_{2n}
\!\left(
\frac{\sqrt{1-q}}{2}\lambda
\right)
=
\sum_{n=0}^{d/2}
w_{2n}^{\,d}(q)\lambda^{2n},
\end{equation}
where the second equality follows from the evenness of $V_s$.  The coefficients
$w_{2n}^{\,d}(q)$ depend on both $q$ and the truncation order $d$, while the
symbol ``$\approx$'' indicates the finite order approximation.  Following the
construction in section~\ref{Recursiveconstructionfrommoments}, we compute the
recursion coefficients using the moment recursion method based on
Eq.~\eqref{recursivigheent} and the recursive
algorithm \eqref{recursivealgorithm}, rather than from the discrete string
equations \eqref{VSRREEQ}, which become cumbersome at high degree. As in the analysis of the asymmetric quartic potential in
section~\ref{Asymmetricquarticpotential}, it is instructive to examine the
continuum string equations.  For the symmetric potential \eqref{potentialdssy},
Eq.~\eqref{polyabv-R} reduces to
\begin{equation}
\sum_{n=1}^{d/2}
n
\binom{2n}{n}
w_{2n}^{\,d}(q)
R(x)^n
=
x.
\label{eq:saddle-R}
\end{equation}
The companion equation \eqref{polyabv-S} is identically satisfied because the
symmetry of the potential implies $S(x)=0$.
\begin{figure}[!htbp]
\centering

\subfigure[$q=0.3$]{
  \includegraphics[width=0.31\textwidth]{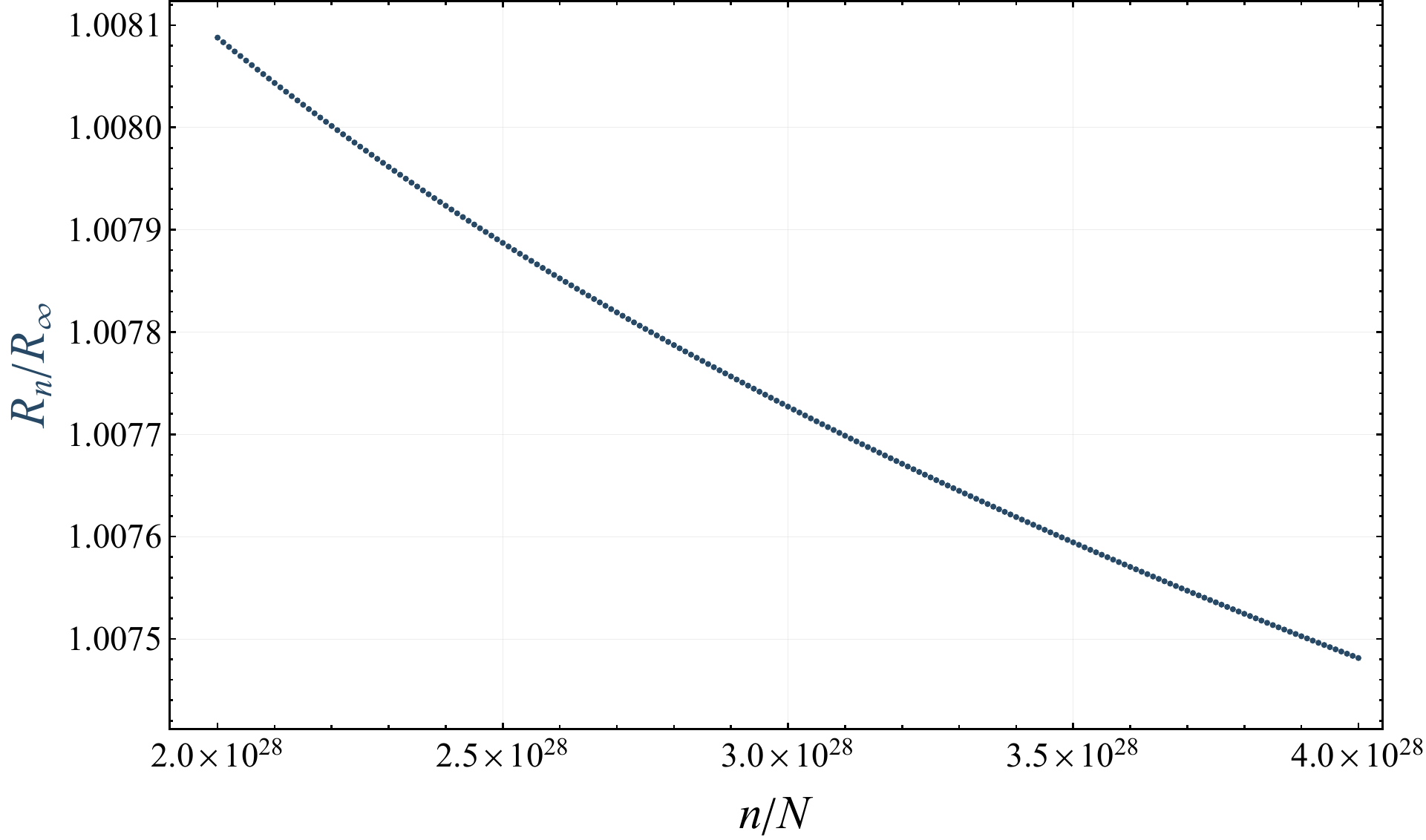}}
\hfill
\subfigure[$q=0.6$]{
  \includegraphics[width=0.31\textwidth]{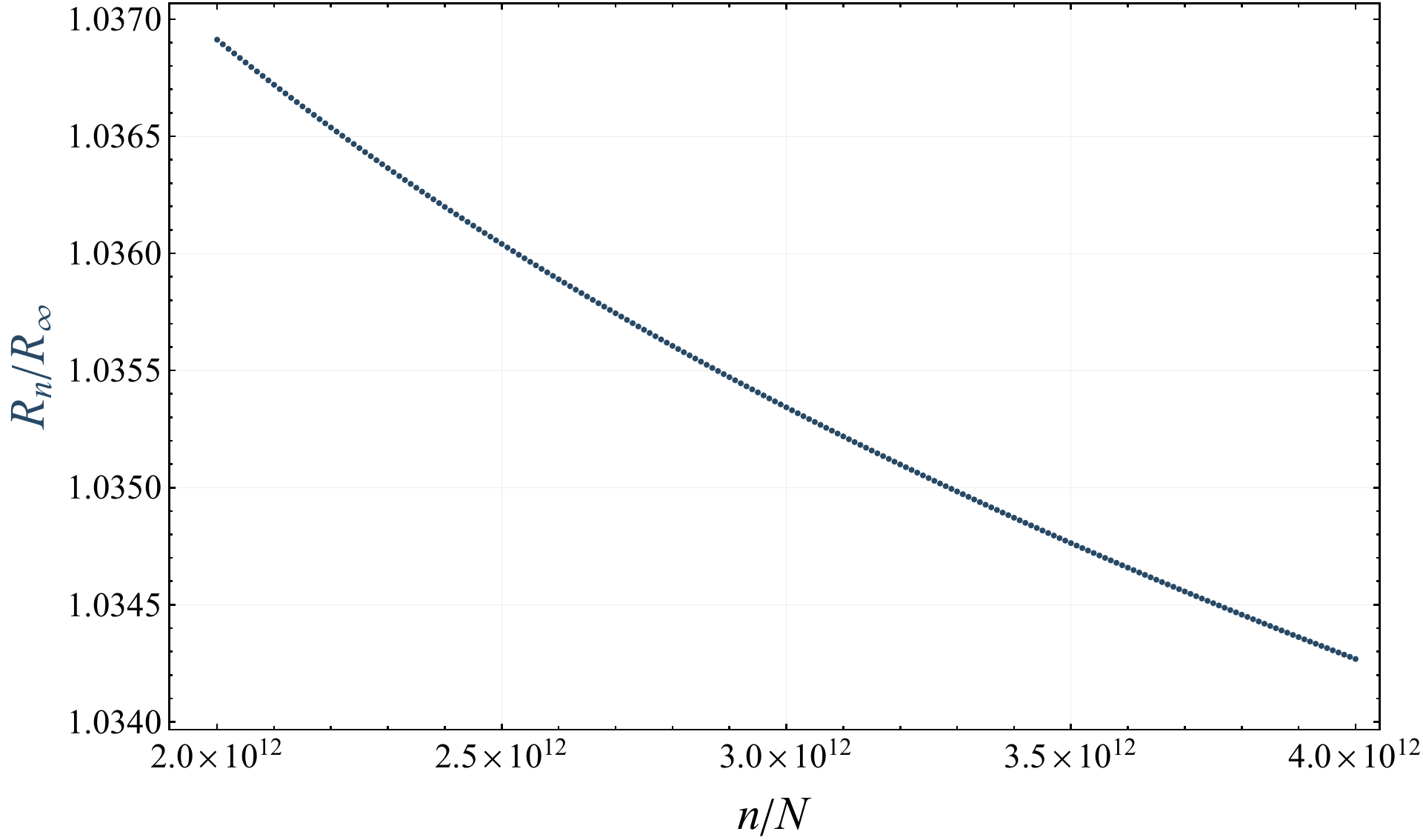}}
\hfill
\subfigure[$q=0.9$]{
  \includegraphics[width=0.31\textwidth]{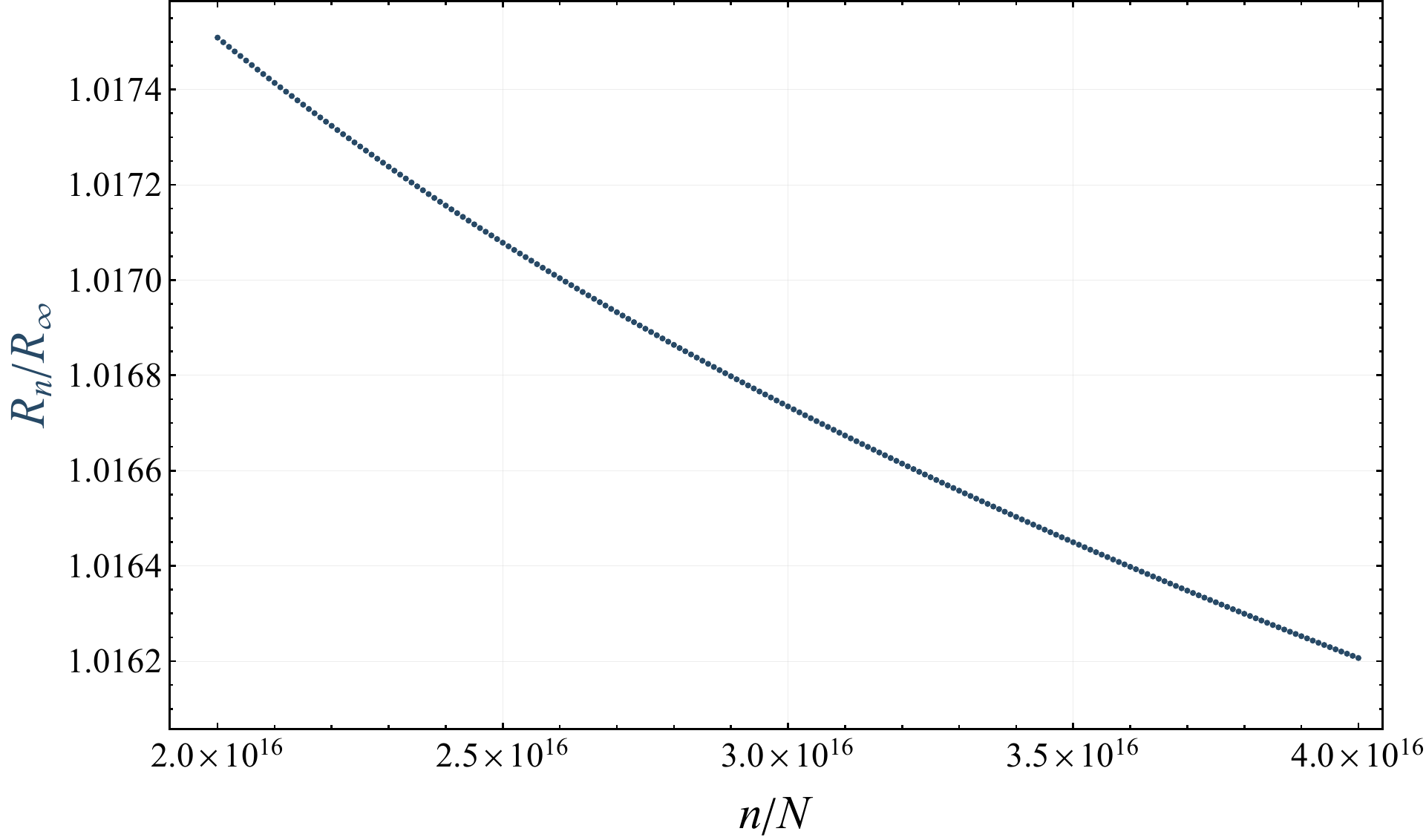}}
\caption{Numerical verification of the asymptotic behavior predicted by
Eq.~\eqref{eq:lanczos-asymptoticsdssy}.  The ratio $R_n/R_\infty$ is shown for
representative values of $q$.}
\label{fig:dssyk-ratio}
\end{figure}
Following the strategy used for the asymmetric quartic potential, we locate the gradient catastrophes by identifying the turning points of $x(R)$,
which satisfy
\begin{equation}
\frac{dx}{dR}
=
\sum_{n=1}^{d/2}
n^2
\binom{2n}{n}
w_{2n}^{\,d}(q)
R(x)^{\,n-1}
=
0.
\end{equation}
For the numerical analysis below, we choose the truncation order $d=18$.  
\begin{figure}[!htbp]
\centering
\subfigure[$N=20, q=0.52$]{
  \includegraphics[width=0.45\textwidth]{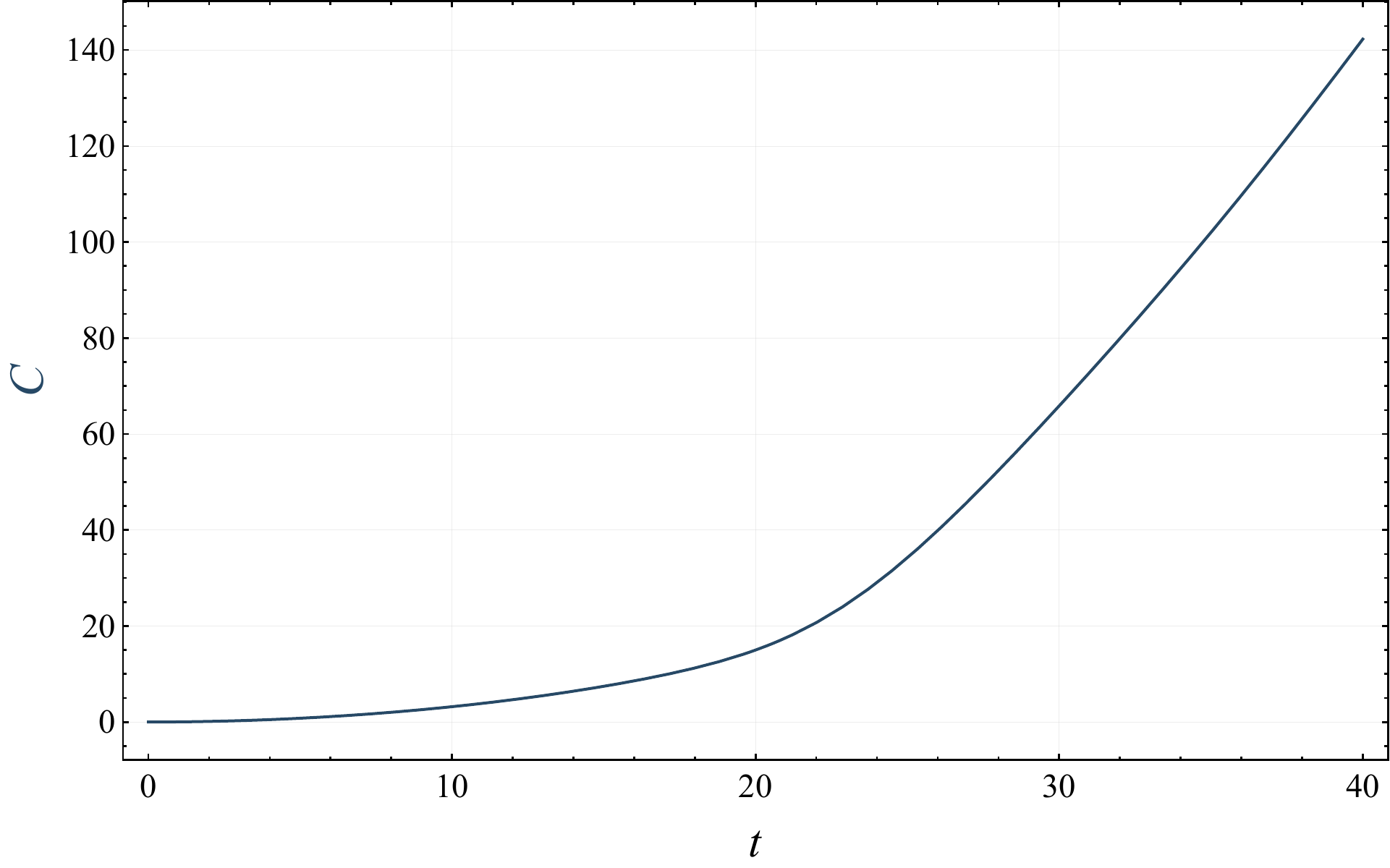}}
\hfill
\subfigure[$N=3\times10^{-6}, q=0.31$]{
  \includegraphics[width=0.45\textwidth]{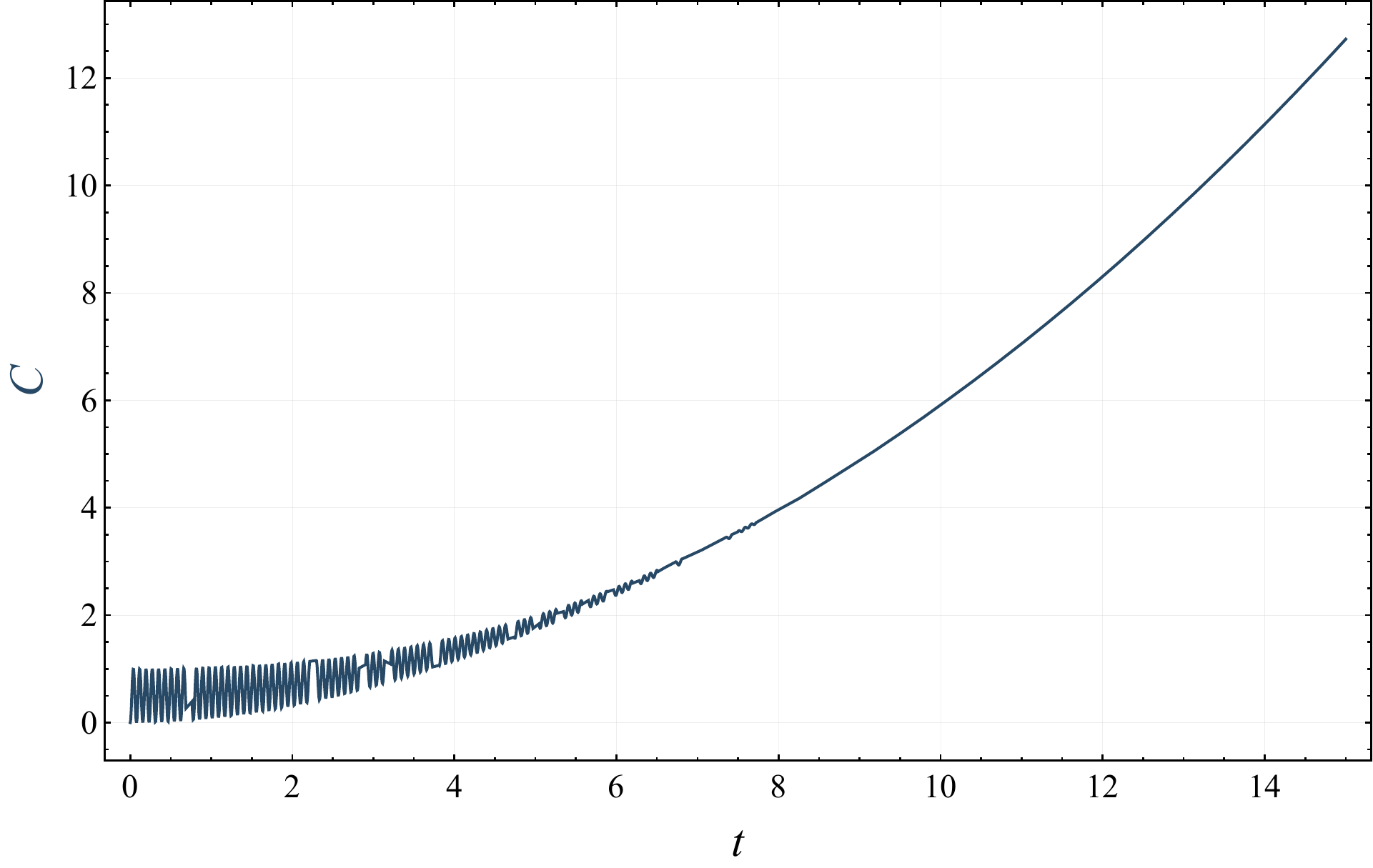}}
\caption{Spread complexity for the DSSYK potential $V_s$.  For
$N=20$ and $q=0.52$, the recursion coefficients contain a transition region,
as shown in Fig.~\ref{fig:dssyk-folds}, while the complexity remains smooth and
increases monotonically.  For $N=3\times10^{-6}$ and $q=0.31$, the two branch
structure of the recursion coefficients produces early time oscillations
followed by monotonic growth.}
\label{fig:dssyk-krylov-complexity}
\end{figure}
The discriminant of the polynomial above vanishes at
\begin{align}\label{zerosofthediscriminant}
q_1 &\approx 0.311527021, &
q_2 &\approx 0.520737446, &
q_3 &\approx 0.632491566, &
q_4 &\approx 0.696692604,
\notag\\
q_5 &\approx 0.758207256, &
q_6 &\approx 0.819373765, &
q_7 &\approx 0.880474263, &
q_8 &\approx 0.942730258.
\end{align}
These critical values partition the interval $0<q<1$ into regions with
different numbers of gradient catastrophes. Figure~\ref{fig:dssyk-folds} compares $R_n$ with $R(x)$ across the regions
separated by the critical values in \eqref{zerosofthediscriminant}. The recursion coefficients can exhibit multiple transition regions, while the recursion function remains accurate not only at small and large $n/N$, but also throughout the smooth intervals between successive transition regions. For $0<x<1$, the recursion functions coincide exactly with the average Lanczos coefficients obtained in Refs.~\cite{Balasubramanian:2024lqk,Nandy:2024zcd}, up to the reversal of the continuum coordinate $x\to1-x$. We also verify the general asymptotic behavior derived in
section~\ref{asymptotics} for the DSSYK potential $V_s$. At truncation order
$d=18$, Eq.~\eqref{eq:lanczos-asymptotics} becomes
\begin{equation}\label{eq:lanczos-asymptoticsdssy}
R_\infty(n)
=
\frac{n^{1/9}}
{24310^{1/9}\left[Nq^{36}(1-q)^9(1+q^9)\right]^{1/9}}.
\end{equation}
Figure~\ref{fig:dssyk-ratio} confirms that $R_n/R_\infty$ approaches one as
$n/N\to\infty$. The spread complexity is obtained from the Schr\"odinger
equation~\eqref{psieq} and displays the same qualitative distinction found for
the quartic potential.
Figure~\ref{fig:dssyk-krylov-complexity} shows that the transition region at
$q=0.52$ does not alter the smooth, monotonic growth of the complexity \footnote{Note that $N$ here denotes only the parameter appearing in the weight function of the orthogonal polynomials \eqref{innerproduct}, and therefore need not be an integer.}.  By
contrast, the two branch structure at $q=0.31$ produces early time oscillations
before the complexity crosses over to monotonic growth.

\section{Conclusions and outlook}\label{Conclusion}
In this work, we have developed an efficient method for computing the recursion
coefficients of high degree polynomial potentials by combining the
moment recursion method with the recursive algorithm.  This approach avoids the
increasingly cumbersome discrete string equations and extends the practical
computation of recursion coefficients to higher degree potentials.  We have
also derived the large-$n$ asymptotic behavior of the recursion coefficients
for general asymmetric potentials.  For $Nw_d=1$, the leading asymptotic form
of $R_n$ reduces to Freud's conjecture.

We have applied these methods to two specific models.  In both cases, the recursion functions capture the overall qualitative behavior of the
recursion coefficients, while gradient catastrophes of the recursion functions
are associated with the appearance of ``chaotic'' transition regions and
provide a natural classification of the parameter space.  For the asymmetric
quartic potential, the discrete string equations show that both $R_n$ and
$S_n$ can develop such regions.  For the DSSYK model, the combination of the
moment recursion method and the recursive algorithm reveals multiple
transition regions in $R_n$, with the recursion function continuing to describe
the smooth intervals between successive regions.  We have also computed the
associated spread complexity in both models.  Our results indicate that a
transition region in the recursion coefficients does not by itself
qualitatively alter the spread complexity, whereas a two branch structure
produces early time oscillations followed by monotonic growth.

Despite extensive studies of transition regions, their detailed structure and
extent remain difficult to characterize analytically.  In particular, deriving
their onset and width from first principles remains an open problem.
Furthermore, double-scaled random matrix models describe two dimensional
quantum gravity~\cite{Banks:1989df,Seiberg:2004at}, in which recursion
coefficients and their string equations play a central role.  Recent
developments in double-scaled matrix models and their connection to
two dimensional quantum gravity are discussed in
Refs.~\cite{Saad:2019lba,Johnson:2019eik,Johnson:2020heh,Johnson:2020exp,
Johnson:2022wsr}.  Against this backdrop, an important direction for future
work is to identify a holographic dual of the spread complexity studied
here~\cite{Caputa:2024sux} and clarify its relation to existing proposals
for holographic complexity~\cite{Susskind:2014rva,Stanford:2014jda,
Brown:2015bva,Brown:2015lvg,Cai:2016xho,Guo:2017rul,Pedraza:2021mkh,Pedraza:2021fgp,Belin:2021bga,Belin:2022xmt,Pedraza:2022dqi,
Carrasco:2023fcj,Jorstad:2023kmq,Jiang:2023jti,Caceres:2023ziv,
Myers:2024vve,Arean:2024pzo,Jiang:2025qai,Miyaji:2025jxy,
Caceres:2025myu,Caceres:2025ypk,Fatemiabhari:2025cyy,
Fatemiabhari:2025usn,Fatemiabhari:2025poq,Nunez:2026vhw}.

\acknowledgments
We are pleased to thank Bowen Chen, Yichao Fu, and Ming-Xuan Liu for their collaboration during the initial stages of this project and for many helpful discussions. We thank Pawel Caputa, Ben Craps, Yu-Xiao Liu and Shan-Ming Ruan for valuable correspondence. JFP is supported by the ‘Atracción de Talento’ program of the Comunidad de Madrid under grant 2020-T1/TIC-20495. LCQ acknowledges support from the Chinese Scholarship Council (CSC) through a graduate scholarship.  Both authors also acknowledge support from the Spanish Agencia Estatal de Investigación through grants CEX2025-001574-S, PID2021-123017NB-I00 and PID2024-156043NB-I00, funded by MCIN/AEI/10.13039/501100011033, and ERDF, EU.

\appendix


\bibliography{Refs}
\bibliographystyle{JHEP}

\end{document}